\documentstyle[prd,aps,twocolumn,epsf,floats,amsfonts,amssymb,amsmath]{revtex}

\newcommand{\bea}{\begin{eqnarray}}\newcommand{\eea}{\end{eqnarray}}
\newcommand{\ba}{\begin{array}}\newcommand{\ea}{\end{array}}
\newcommand{\bit}{\begin{itemize}}\newcommand{\eit}{\end{itemize}}
\newcommand{\ben}{\begin{enumerate}}\newcommand{\een}{\end{enumerate}}

\newcommand{\lab}{\label}

\newcommand{\lf}{\left}
\newcommand{\noi}{\noindent}
\newcommand{\non}{\nonumber}
\newcommand{\pa}{\partial}\newcommand{\ran}{\rangle}

\newcommand{\ri}{\right}

\newcommand{\ga}{\gamma}\newcommand{\Ga}{\Gamma}

\newcommand{\ka}{\kappa}

\newcommand{\Om}{\Omega}\newcommand{\I}{{_I}}

\newcommand{\TA}{{\textsf{A}}}
\newcommand{\TB}{{\textsf{B}}}
\newcommand{\TR}{{_T}}
\begin{document}

\twocolumn[\hsize\textwidth\columnwidth\hsize\csname@twocolumnfalse\endcsname

\typeout{--- Title page start ---}

\begin{flushright}
Imperial/TP/0-01/11 \\
{\tt quant-ph/0102128}\\
\end{flushright}
\vskip 1cm

\title{Bateman's dual system revisited: I. Quantization, geometric phase and
relation with the ground--state
energy of the linear harmonic oscillator}
\author{Massimo Blasone${}^{\sharp \flat}$ and Petr Jizba${}^{\sharp
\natural}$}

\address{ $ $ \\[2mm]
${}^{\sharp}$   Blackett  Laboratory, Imperial  College,   Prince
Consort Road, London  SW7 2BZ, U.K.
\\ [1mm] ${}^{\natural}$ Institute of Physics, University of Tsukuba,
Ibaraki 305-8571, Japan
\\ [1mm] ${}^{\flat }$
Dipartimento di  Fisica and INFN, Universit\`a di Salerno, I-84100 Salerno,
Italy
\\[3mm] E-mails:  m.blasone@ic.ac.uk, p.jizba@ic.ac.uk}

\vspace{4mm}

\date{{\bf Version}, \today}

\maketitle

\begin{center}
{\small \bf Abstract}
\end{center}
\begin{abstract}
\noindent
By using the Feynman--Hibbs prescription for the evolution
 amplitude,
we quantize the system of a  damped harmonic oscillator coupled
to its time--reversed image, known as Bateman's dual system.
 The time--dependent quantum states of such a system
are constructed and
discussed entirely in the framework of the classical theory.
The corresponding geometric (Pancharatnam) phase is calculated
and found to be directly related
 to the ground--state energy of the 1D linear harmonic
oscillator to which the 2D system reduces  under
appropriate constraint.
\\

\vspace{3mm}
\draft
\end{abstract}
\vskip8mm]



\section{Introduction}

\noindent From the very outset of quantum theory a tremendous effort
has been devoted to answering the question where is the boundary
between the classical (macroscopic) and the quantum (microscopic)
world, or in other words, when a quantum system starts to behave
classically. The correspondence principle introduced by Bohr in the end
of 20's served as a heuristic prescription to construct quantum
mechanics. Roughly speaking, the quantum theory should approach the
classical theory in the limit of large quantum numbers. However, in
general, this limit is quite subtle \cite{AP1,RO1}.  On the other hand,
a typical statement in majority of standard textbooks is  that
classical mechanics applies in the limit $\hbar \rightarrow 0$, and
this paradigm has become over the years a starting point for a host of
semi--classical asymptotic treatments\cite{VPM1,RGL2,AMO1}. Yet, still
many precautions should be taken when it comes to non--physical
infinities at caustics, boundary layers analysis  or to connection
rules\cite{AT,RGL1}.  In terms of canonically conjugated variables, say
$p$ and $q$, the connection between the classical and quantum
descriptions has been established by Ehrenfest's theorems
\cite{AP1,RGL2,AM} which give the law of motion for the mean values of
the operators ${\hat{p}}$ and ${\hat{q}}$ in a form emulating the
classical equations of motion. However, only for systems with quadratic
Hamiltonians is this correspondence exact, i.e. the mean values
$\langle {\hat{p}} \rangle$ and $\langle {\hat{q}} \rangle$ follow
classical phase--space trajectories\cite{RGL2}. In particular, the
system of quantum harmonic oscillators with bilinear coupling can be
completely characterized using classical trajectories because the
quantum Wigner function satisfies the classical Liouville equation
\cite{EW1}.

\vspace{3mm}

\noindent In the present article, the first of the series,
we attempt to shed some further
light on the quantum--classical relation by studying the
quantization of the  2D system of a
damped harmonic oscillator coupled to its time--reversed
image, first introduced by Bateman\cite{BAT}.
Bateman's dual system received
considerable attention in the past
as it represents a simple explicit example of a
dissipative system which could be tackled by means of canonical
quantization\cite{HD1,HF1,GV1}. However, the Quantum Mechanics (QM) of this
system is plagued with many
conceptual problems\cite{HD1,HF1} (e.g., the wave functions cannot be
normalized  in the usual manner, the Hamiltonian is not self--adjoint
and represents the energy only for a restricted set of dynamical
solutions) and  it was shown\cite{GV1}  that
a consistent quantization can only be achieved in Quantum Field
Theory (QFT).
More recently, Bateman's system has been studied in connection
with (Chern-Simons) gauge theories\cite{MB1},
as an example of an exactly decoherent system\cite{halliwell}
and as a toy model for the recent proposal by
G.'t Hooft about deterministic QM\cite{TH1,BJV1}.
These aspects as well as the QFT of the Bateman
system will be the object of future papers. The aim of this paper is to
present a thorough analysis of the QM of this system which will be the
basis for the next papers: our results  include the
time--dependent wave functions and the geometric phases
associated to them. We also study the reduction of the 2D Bateman
system to the 1D linear harmonic oscillator (l.h.o.).

\vspace{3mm}
\noi An  outline of the paper  is as follows:

\vspace{3mm}

\noindent
In Section II we quantize Bateman's system by using
 the Feynman--Hibbs prescription for the
time--evolution amplitude (kernel)\cite{RPF1,HK1,DYS1}: this
allows us to avoid the pitfalls of canonical quantization.
We show  that the kernel is  fully expressible in terms of
solutions of the classical equations of motion and that it is
invariant with respect to the choice  of the fundamental system of
those solutions. This might be viewed as a
two--dimensional extension of an existing one--dimensional
result\cite{DYS1}. An important ingredient of the kernel calculation
is the fluctuation factor\cite{HK1}. This is calculated by
employing  the Van Vleck--Pauli--Morette
determinant technique  which allows us to avoid a direct
manipulation of the Schr{\"o}dinger  equation. Instead, only the
phase--space structure of the classical  solutions (e.g., Lagrangian
manifold) is used.

\vspace{3mm}

\noindent In Section III we use
Mehler's formula \cite{GR1,MF1} for a
spectral decomposition of the kernel in order
to obtain the time--dependent
wave functions in hyperbolic  radial
coordinates ($r,u$). They are expressed in terms of
generalized Laguerre polynomials and are shown to  satisfy the
correct time--dependent Schr{\"o}dinger equation. The explicit form of the wave
functions uncovers the root of the difficulties connected with the
canonical quantization - the unboundedness of Bateman's system in the
hyperbolic angle $u$. Use of the  radial kernel\cite{HK1} allows us to
factorize away the explicit $u$ dependence.  The ``radial'' wave
functions then correctly fulfil both orthonormalization and
completeness relations. In addition, the radial wave functions
satisfy the radial, time--dependent Schr{\"o}dinger equation with
the Hamiltonian ${\hat{H}}_l$.  For the ``azimuthal'' quantum number
$l= \pm\frac{1}{2}$, the latter turns out to be formally identical
with the Hamiltonian of the  1D l.h.o..

\vspace{3mm}

\noindent To better understand the structure of the wave functions,
we focus our attention on the algebraic setting of Bateman's
Hamiltonian. Identifying the dynamical group as $SU(1,1)$, we are able to
 pinpoint the structure of the ground state, which turns
out to be a (squeezed) coherent state.  The aforementioned peculiar
behavior of the wave functions is  then attributed to the remarkable
properties of the $SU(1,1)$ group representations:  the
unboundedness of Bateman's system in the variable $u$ could be seen as a
consequence of the non--existence of a unitary irreducible
representation of $SU(1,1)$ in which the generator $J_2$ would have  at
the same time a real and discrete spectrum\cite{GL1}.
The requirement of discreteness
of the $J_2$ spectrum then leads to an {\em effective} non--hermiticity and
oddness of $J_2$ under time reversal. To accommodate this point in the
Feynman--Hibbs kernel  prescription, a new inner product has to be
defined. In fact, one of the merits of the presented method is
that it naturally provides the consistent inner product for  wave functions
(the rather artificial and involved method of Racah\cite{HD1}
is not needed).  In the case when we restrict our attention to the
stationary  quantum states, the connection with  existing canonical
quantization results\cite{HD1,HF1,GV1} is readily established.

\vspace{3mm}

\noindent With the (full) time--dependent wave functions at hand, we
are able to calculate in Section IV  the exact geometric
(Pancharatnam) phase for
Bateman's dual system. We find
that Pancharatnam's phase is explicitly $\hbar$ independent
and consists of three autonomous contributions:
overall ground--state fluctuations of $\hat{p}$ and $\hat{x}$ gathered during
the period of evolution and the Morse index.

\vspace{3mm}

\noindent We then show that the (full) wave
functions become periodic in configuration space when
the hyperbolic angle $u$ solves the classical equations of  motion.
In this case the period of the wave functions matches the inverse of the
reduced  frequency of the original Bateman dual system and
Pancharatnam's phase boils down to the ordinary Berry--Anandan phase.
In Section V we take this observation over to the radial wave
functions  and then, by setting $l=\pm\frac{1}{2}$, to the 1D l.h.o..
Because the harmonic oscillator wave functions obtained in
this way are constructed entirely from the  fundamental system of
solutions of Bateman's dual system, the Berry--Anandan phase bears an
imprint (or memory) of the original 2D system even after the reduction to
the 1D l.h.o. is performed.  The geometric phase thus
obtained can be directly identified with the zero point energy of
the 1D l.h.o., and,  in general,  it is different
from the usual $E_{0} = \hbar \Omega /2$. This  is in line with
the results obtained in\cite{BJV1}.

\section{Time--evolution amplitude (kernel) for Bateman's dual system}

\noindent In the following  a decisive r{\^o}le will be
played by the matrix elements of the time evolution operator
$U(t_{b},t_{a})$ in the localized state basis
\begin{equation}
\langle {\bf{x}}_{b};t_{b}| {\bf{x}}_{a};t_{a} \rangle \equiv
\langle {\bf{x}}_{b}| U(t_{b},t_{a})| {\bf{x}}_{a} \rangle \, .
\label{ker01}
\end{equation}
\noindent They are referred as {\em time--evolution amplitudes},  or
simply {\em kernels}. Due to the fact that  $U(t_{b},t_{a})$ fulfills
the time dependent Schr{\"odinger} equations
\begin{eqnarray}
i\hbar \frac{\partial}{\partial t_b} U(t_b,t_a) &=& {\hat{H}}\,
U(t_b,t_a)\, , \nonumber \\ i\hbar \frac{\partial}{\partial t_b}
U(t_a,t_b) &=& -\, U(t_a,t_b)\, {\hat{H}} \qquad, \;
t_b >t_a\, , \label{ker02}
\end{eqnarray}
\noindent the kernel satisfies  the equations \cite{SCH1}:
\begin{eqnarray}
&&i\hbar \frac{\partial}{\partial t_b} \langle {\bf{x}}_b;t_b|
{\bf{x}}_a;t_a \rangle =  {\hat{H}}\left(-i\hbar\,\partial_{{\bf{x}}_b
},  {\bf{x}}_b \right) \,   \langle {\bf{x}}_b;t_b| {\bf{x}}_a;t_a
\rangle \, ,\nonumber \\  &&i\hbar \frac{\partial}{\partial t_b}
\langle {\bf{x}}_a;t_a|  {\bf{x}}_b;t_b \rangle
\label{ker03}
\\ \non
&&= -\, {\cal{T}}{\hat{H}}^{\dagger}
\left(-i\hbar\,
\partial_{{\bf{x}}_b},  {\bf{x}}_b \right) {\cal{T}}^{-1}\,  \langle
{\bf{x}}_a;t_a|  {\bf{x}}_b;t_b \rangle
\quad , \; t_b > t_a\; ,
\end{eqnarray}
\noindent with the initial condition
\begin{equation}
\lim_{t_b \rightarrow t_a} \langle {\bf{x}}_a ;t_a| {\bf{x}}_b; t_b
\rangle = \delta({\bf{x}}_a -{\bf{x}}_b) \, .
\end{equation}
\noindent Here ${\cal{T}}$ is the (anti--unitary) time reversal
operator and ${\hat{H}}^{\dagger}$ is the Hermitian--adjoint
Hamiltonian  (in most applications ${\hat{H}}$ is both Hermitian and
even under time
reversal so
$\dagger$ and ${\cal{T}}$ are  usually omitted). An
important observation is that  for
quadratic Hamiltonians the kernel has a very simple form, namely
\begin{equation}
\langle {\bf x}_{b};t_{b}| {\bf x}_{a};t_{a} \rangle =
F[t_{a},t_{b}]\, \mbox{exp}\left( \frac{i}{\hbar}S_{cl}[{\bf x}]
\right)\, ,\;\;\;\; t_{b}>t_{a} \, .
\label{ker05}
\end{equation}
\noindent The function $F[t_{a},t_{b}]$ is the so called {\em
fluctuation factor} \cite{HK1} and is independent of both ${\bf
x}_{a}$ and ${\bf x}_{b}$ \cite{RPF1,HK1}. The form (\ref{ker05}) is
usually attributed to Feynman and Hibbs\cite{HK1,DYS1}, but one may
readily see that it  is nothing but the kernel version of
the celebrated WKB approximation (often referred  to as the Van Vleck
\cite{RGL1} formula), which turns out to be
an exact relation for quadratic Hamiltonians.

\vspace{3mm}

\noindent For a system with a time--independent Hamiltonian, the kernel
reads
\begin{equation}
\langle {\bf{x}}_b;t_b|{\bf{x}}_a;t_a \rangle = \langle {\bf{x}}_b|
\mbox{exp}\left(-\frac{i}{\hbar}\, {\hat{H}}(t_b - t_a) \right)|
{\bf{x}}_a \rangle \, .
\label{ker06}
\end{equation}
\noindent Inserting the resolution of unity (completeness relation)
\begin{displaymath}
\sum_{m} |\psi_m \rangle\langle \psi_m| = 1\, .
\end{displaymath}
\noindent (here $|\psi_m\rangle$ are orthonormal base kets at $t=0$
spanning the Hilbert space) into (\ref{ker06}), we obtain that
\begin{equation}
\langle {\bf{x}}_b;t_b|{\bf{x}}_a;t_a \rangle = \sum_m
\psi_m({\bf{x}}_b,t_b) \psi_m^*({\bf{x}}_a,t_a)\, .
\label{ker7}
\end{equation}
\noindent Here we have identified $\psi_m({\bf{x}},t) = \langle
{\bf{x}}|\psi_m(t)\rangle$. The symbol $*$ denotes usual complex
conjugation.  Note that $\psi_m({\bf{x}},t)$ and
$\psi_m^*({\bf{x}},t)$, obey first and second equation in
(\ref{ker03}), respectively.

\vspace{3mm}

\noindent In the following we shall use  the Feynman--Hibbs
prescription (\ref{ker05})  for  the the quantization of Bateman's dual
system\cite{HD1,GV1}. Our analysis will also reveal
a host of subtleties which are hidden in the seemingly clear relation
(\ref{ker05}).

\subsection{Lagrangian and classical equations of motion}
\noindent Bateman's dual model describes a 2D interacting system of
damped--amplified harmonic oscillators. The corresponding Lagrangian
reads\cite{BAT,HD1,HF1,GV1,MB1,MF1}
\begin{eqnarray}
L &=& m\dot{x}\dot{y} + \frac{\gamma}{2}(x \dot{y} - \dot{x} y) -
\kappa xy
\label{lag1}
\end{eqnarray}
giving the (classical) equations of motion:
\begin{eqnarray} \non
m{\ddot{x}}_{cl} + \gamma {\dot{x}}_{cl} + \kappa x_{cl} &=& 0\, , \\
m{\ddot{y}}_{cl} - \gamma {\dot{y}}_{cl} + \kappa y_{cl} &=& 0 \, ,
\end{eqnarray}
\noindent It is interesting to observe that the equation for $x$
describes the damped harmonic oscillator, while the equation for $y$
characterizes the amplified oscillator. In addition, with appropriate
initial conditions both systems  are mutual mirror images. In this
sense it may be sometimes helpful  to think of $y$ as describing an
effective degree of freedom for the reservoir to which system with the
$x$ degree of freedom is coupled\cite{HD1,GV1,MB1}.

\vspace{3mm}

\noi In the following it will be useful to work with the rotated
variables\cite{MB1}:
$x_1 = (x+y)/\sqrt{2}$, $x_2=(x-y)/\sqrt{2}$. Then

\begin{eqnarray}
L &=& \frac{m}{2}(\dot{x}_{1}^{2} - \dot{x}_{2}^{2}) +
\frac{\gamma}{2} (\dot{x}_{1}x_{2} - \dot{x}_{2}x_{1}) -
\frac{\kappa}{2}(x_{1}^{2} - x_{2}^{2})\nonumber \\ &=&
\frac{m}{2}\dot{\bf{x}}\dot{\bf{x}} + \frac{\gamma}{2}\, {\bf{x}}
\wedge \dot{\bf{x}} - \frac{\kappa}{2} {\bf{x}}{\bf{x}} \, .
\label{lag2}
\end{eqnarray}

\noindent where we introduced the notation ${\bf{a}}{\bf{b}} = g_{\alpha
\beta}\, a^{\alpha} b^{\beta}$, ${\bf{a}}\wedge {\bf{b}} =
\varepsilon^{\alpha \beta}a_{\alpha}b_{\beta}$ and
$x^{\alpha}=(x_1,x_2)$
with the metric tensor $g_{\alpha \beta} = (\sigma_{3})_{\alpha
\beta}$ (note also that $\varepsilon^{\alpha \beta} =
-\varepsilon_{\alpha \beta}$).
The corresponding conjugate momenta read
\begin{equation}
{\bf p} = m {\dot {\bf x}} - \frac{1}{2}\gamma \sigma_{1} {\bf x}\, .
\label{cm12}
\end{equation}

\noi In $(x_{1},x_{2})$ coordinates the equations of motion read
\begin{equation}
m\, \ddot{\bf{x}}_{cl} + \gamma \sigma_{1}\, \dot{\bf{x}}_{cl} +
\kappa \, {\bf{x}}_{cl} = {\bf{0}}\,
\label{e2}
\end{equation}

\noindent Notice that if ${\bf u}(t)$ is a solution of (\ref{e2}) so
are $\sigma_{1}{\bf u}(t)$, $\sigma_{3}{\bf u}(-t)$ and
$i\sigma_{2}{\bf u}(-t)$. For the future reference it is useful to
realize that the Wronskian is $t$ independent (i.e. it is a time
invariant of the system). Indeed, in our case the Wronskian has the
form ($t_{0}$ is arbitrary):
\begin{equation}
W(t) = W(t_{0}) \mbox{exp}\left(-\int_{t_{0}}^{t}dt \,
\mbox{Tr}\left(\frac{\gamma}{m}\sigma_{1}\right)\right)\, ,
\label{w2}
\end{equation}

\noindent  Eq.(\ref{w2})
is nothing but Liouville's theorem of a differential calculus applied
to (\ref{e2}).

\subsection{Classical action}

\noindent Using the usual definition for the action:

\begin{displaymath}
S[{\bf{x}}] = \int_{t_{a}}^{t_{b}} dt\, L \, ,
\end{displaymath}

\noindent we can write

\bea &&S_{cl}[{\bf{x}}] = \int_{t_{a}}^{t_{b}}dt\,
\left[\frac{m}{2}\left(\frac{d}{dt}(x_1\dot{x}_1  -x_2\dot{x}_2)
- x_1\ddot{x}_1 + x_2\ddot{x}_2 \right)\right.\non
\\  &&
\mbox{\hspace{10mm}}
\left. -\frac{\gamma}{2}(x_1\dot{x}_2 - x_2\dot{x}_1 )
- \frac{\kappa}{2}(x_1^2 - x_2^2) \right] \non
\\ &&= \frac{m}{2}(x_1\dot{x}_1 -
x_2\dot{x}_2)|_{t_{a}}^{t_{b}}
- \int_{t_{a}}^{t_{b}}dt\,
\frac{\bf{x}}{2}(m\ddot{\bf x}+ \gamma \sigma_{1}\dot{\bf  x} + \kappa
{\bf x} )  \non
\\ &&=
\frac{m}{2}[{\bf{x}}_{cl}(t_{b})\dot{\bf{x}}_{cl}(t_{b}) -
{\bf{x}}_{cl}(t_{a})\dot{\bf{x}}_{cl}(t_{a}) ]\, .
\label{e7}
\eea

\subsection{Fundamental system of solutions}

\noindent A fundamental system of solutions (i.e. a maximal system of
linearly independent  solutions) for  Eq.(\ref{e2})
consists of {\em four} real $1\times2$ vectors ${\bf u}_{i}$
$(i=1,2,3,4)$. The reason why there are four independent solutions
is, roughly speaking, a result of the fact that we have two boundary
conditions for each index. Independence of solutions may be checked
via the Wronskian, which has to be non--zero at least at one time
$t$ (actually the Wronskian is time independent here). In our case the
Wronskian is the determinant of a $4 \times 4$ matrix:
\bea W(t) = W(t_{0}) = \left|
\begin{array}{cccc}
{\bf u}_{1} & {\bf u}_{2} & {\bf u}_{3} & {\bf u}_{4}\\ \dot{\bf
u}_{1} & \dot{\bf u}_{2} & \dot{\bf u}_{3} & \dot{\bf u}_{4}
\end{array} \right|\, .
\eea
An important technical simplification may be achieved by
realizing
that we may always find such a fundamental
system where  two arbitrary solutions
(say, ${\bf u}_{3}$ and ${\bf u}_{4}$) are set to zero at
$t_{a}$.  This is due to the fact that in order to fulfill the
boundary condition on ${\bf x}(t_{a})$ we need only two
linearly independent vectors.
Let us fix the following
convention: ${\bf u}_{3} \equiv {\bf v}_{1}$ and ${\bf u}_{4} \equiv
{\bf v}_{2}$.  Then the condition on the fundamental system may be
rephrased as
\begin{displaymath}
W(t) = W(t_{a}) = \left| {\bf u}_{1}(t_{a}) {\bf u}_{2}(t_{a})\right|
\times \left| \dot{\bf v}_{1}(t_{a}) \dot{\bf v}_{2}(t_{a})\right|
\not= 0\, .
\end{displaymath}
Note that if we had assumed the existence of a fundamental
system having three linearly independent
 solutions being zero at $t_{a}$, the Wronskian
would vanish identically. Any real solution of
(\ref{e2}) might thus be written as
\begin{displaymath}
{\bf x}_{cl}(t) = \alpha_{1}{\bf u}_{1}(t) + \alpha_{2}{\bf u}_{2}(t)
+ \beta_{1}{\bf v}_{1}(t) + \beta_{2}{\bf v}_{2}(t) \, ,
\end{displaymath}
with $\alpha_{i}$ and $\beta_{i}$ being real numbers.
Applying Cramer's rule, the solution ${\bf {x}}_{cl}(t)$ with two
fixed points ${\bf x}_{cl}(t_{a}) \equiv {\bf x}_{a}$ and ${\bf
x}_{cl}(t_{b}) \equiv {\bf x}_{b}$ reads

\bea {\bf x}_{cl}(t) &=& \frac{\left[ {\bf u}_{1}(t)\, D_{1} + {\bf
u}_{2}(t)\, D_{2} + {\bf v}_{1}(t)\, D_{3} + {\bf v}_{2}(t)\, D_{4}
\right]}{U_{a}V_{b}}\, , \non \\
\label{cr1}
\eea
\noindent where $U_{a} = \left| {\bf u}_{1}(t_{a}) {\bf u}_{2}(t_{a})
\right|$,  $V_{b} = \left| {\bf v}_{1}(t_{b}) {\bf v}_{2}(t_{b})
\right|$ and
\bea D_{1} &=& \left| \begin{array}{llll} {\bf x}_{a} & {\bf
u}_{2}(t_{a}) & {\bf 0} & {\bf 0} \\ {\bf x}_{b} & {\bf u}_{2}(t_{b})
& {\bf v}_{1}(t_{b}) & {\bf v}_{2}(t_{b})
\end{array}
\right|  = {\bf x}_{a} \wedge {\bf u}_{2}(t_{a}) \times V_{b}\, ,\non
\\ && \non \\ D_{2} &=& \left| \begin{array}{llll} {\bf u}_{1}(t_{a})
& {\bf x}_{a}  & {\bf 0} & {\bf 0} \\ {\bf u}_{1}(t_{b}) & {\bf x}_{b}
& {\bf v}_{1}(t_{b}) & {\bf v}_{2}(t_{b})
\end{array}
\right|  =- {\bf x}_{a} \wedge {\bf u}_{1}(t_{a}) \times V_{b}\, ,\non
\\ && \non \\ D_{3} &=& \left| \begin{array}{llll} {\bf u}_{1}(t_{a})
& {\bf u}_{2}(t_{a}) & {\bf x}_{a} & {\bf 0} \\ {\bf u}_{1}(t_{b}) &
{\bf u}_{2}(t_{b}) & {\bf x}_{b} & {\bf v}_{2}(t_{b})
\end{array}
\right| \, ,\non \\ &&\non \\ D_{4} &=& \left| \begin{array}{llll}
{\bf u}_{1}(t_{a}) & {\bf u}_{2}(t_{a}) & {\bf 0} & {\bf x}_{a} \\
{\bf u}_{1}(t_{b}) & {\bf u}_{2}(t_{b}) & {\bf v}_{1}(t_{b}) & {\bf
x}_{b}
\end{array}
\right| \, .   \eea

\noindent An equivalent, and more useful, way of writing ${\bf
x}_{cl}(t)$ is to expand it in terms of ${\bf x}_{a}$ and ${\bf
x}_{b}$. After some algebra we get
\begin{equation}
{\bf x}_{cl}(t) = \frac{\left[x_{a}^{1}\,{\bf B}_{1}(t) + x_{a}^{2}\,
{\bf B}_{2}(t) + x_{b}^{1}\,{\bf B}_{3}(t) + x_{b}^{2} \, {\bf
B}_{4}(t)\right] }{U_{a} V_{b}}\, ,
\label{eq44}
\end{equation}
\noindent where
\begin{equation}
{\bf B}_{i}(t) = \left( \begin{array}{c} B_{i}^{1}(t)\\ B_{i}^{2}(t)
\end{array}
\right)\, .
\end{equation}
\noindent  $B_{i}^{1}$ and $B_{i}^{2}$ are given by the determinant $D$,
\begin{equation}
D = \left| \begin{array}{llll} {\bf u}_{1}(t_{a}) & {\bf u}_{2}(t_{a})
& {\bf 0} & {\bf 0} \\ {\bf u}_{2}(t_{b}) & {\bf u}_{2}(t_{b}) & {\bf
v}_{1}(t_{b}) & {\bf v}_{2}(t_{b})
\end{array}
\right| = U_{a} V_{b}\, ,
\label{det}
\end{equation}
\noindent with $i$-th row substituted by $(u_{1}^{1}(t),
u_{2}^{1}(t),v_{1}^{1}(t),v_{2}^{1}(t))$ or $(u_{1}^{2}(t),
u_{2}^{2}(t),v_{1}^{2}(t),v_{2}^{2}(t))$ respectively. So for example:
\begin{equation}
B_{3}^{1}(t) = \left| \begin{array}{llll} {\bf u}_{1}(t_{a}) & {\bf
u}_{2}(t_{a}) & {\bf 0} & {\bf 0} \\ u_{1}^{1}(t) & u_{2}^{1}(t) &
v_{1}^{1}(t) & v_{2}^{1}(t)\\ u_{1}^{2}(t_{b}) & u_{2}^{2}(t_{b}) &
v_{1}^{2}(t_{b}) & v_{2}^{2}(t_{b})
\end{array}
\right|\, .
\end{equation}
\noindent As a result, the classical action $S_{cl}[{\bf x}]$ might be
written as
\begin{eqnarray} &&S_{cl}[{\bf x}] = \frac{m}{2D}
\left[  x_{a}^{1}\, {\bf x}_{b}
\dot{\bf B}_{1}(t_{b}) + x_{a}^{2}\, {\bf x}_{b} \dot{\bf
B}_{2}(t_{b}) + x_{b}^{1}\, {\bf x}_{b} \dot{\bf
B}_{3}(t_{b})\right.\non \\ &&\mbox{\hspace{17mm}}\left. + \,
x_{b}^{2}\, {\bf x}_{b} \dot{\bf B}_{4}(t_{b}) - x_{a}^{1}\, {\bf
x}_{a} \dot{\bf B}_{1}(t_{a}) - x_{a}^{2}\, {\bf x}_{a} \dot{\bf
B}_{2}(t_{a})\right.\non \\ &&\mbox{\hspace{17mm}}\left.- \,
x_{b}^{1}\, {\bf x}_{a} \dot{\bf B}_{3}(t_{a}) - x_{b}^{2}\, {\bf
x}_{a} \dot{\bf B}_{4}(t_{a}) \right]\, .
\label{Scl1}
\end{eqnarray}
\noindent An explicit representation of the action
is
\begin{eqnarray*}
&&S_{cl}[{\bf x}] = \frac{m}{2D} \left[ -(x_{a}^{1})^{2}\,
\dot{B}_{1}^{1}(t_{a}) +(x_{a}^{2})^{2}\, \dot{B}_{2}^{2}(t_{a})
\right.
\non \\ &&
\mbox{\hspace{2mm}}\left. +(x_{b}^{1})^{2}\, \dot{B}_{3}^{1}(t_{b})
-(x_{b}^{2})^{2}\,
\dot{B}_{4}^{2}(t_{b})\right. \non \\ && \mbox{\hspace{2mm}}\left. +
x_{a}^{1}x_{a}^{2}\, \left(\dot{B}_{1}^{2}(t_{a}) -
\dot{B}_{2}^{1}(t_{a}) \right) + x_{b}^{1}x_{b}^{2}\,
\left(\dot{B}_{4}^{1}(t_{b}) - \dot{B}_{3}^{2}(t_{b})
\right)\right. \non \\ && \mbox{\hspace{2mm}}\left.+
x_{a}^{1}x_{b}^{1}\, \left(\dot{B}_{1}^{1}(t_{b}) -
\dot{B}_{3}^{1}(t_{a}) \right) - x_{a}^{1}x_{b}^{2}\,
\left(\dot{B}_{1}^{2}(t_{b}) + \dot{B}_{4}^{1}(t_{a})
\right)\right. \non \\ && \mbox{\hspace{2mm}}\left.+
x_{a}^{2}x_{b}^{2}\, \left(\dot{B}_{4}^{2}(t_{a}) -
\dot{B}_{2}^{2}(t_{b}) \right) + x_{a}^{2}x_{b}^{1}\,
\left(\dot{B}_{2}^{1}(t_{b}) + \dot{B}_{3}^{2}(t_{a}) \right)
\right]\, .  \,
\end{eqnarray*}
\noindent Using the basic properties of determinants  it is possible to
show now
that  both $S_{cl}[{\bf x}]$ and ${\bf x}_{cl}(t)$ are independent of
the choice of the fundamental system of solutions.
We show this in Appendix B.

\vspace{3mm}

\subsection{Fluctuation factor}

\noindent We can now take advantage of the Feynman--Hibbs
observation\cite{RPF1,HK1} about the time--evolution amplitude
(kernel) for systems governed by quadratic Hamiltonians:
\begin{equation}
\langle {\bf x}_{b};t_{b}| {\bf x}_{a};t_{a} \rangle = F[t_{a},t_{b}]
\, \mbox{exp}\left( \frac{i}{\hbar}S_{cl}[{\bf x}] \right)\, ,\;\;\;\;
t_{b}>t_{a} \, .
\label{ker111}
\end{equation}
\noindent As remarked, the fluctuation factor $F[t_{a},t_{b}]$ is
independent of ${\bf x}_{a}$ and ${\bf x}_{b}$. In addition, from
(\ref{ker111}) follows that
\begin{displaymath}
F[t_{a},t_{b}] = \langle 0; t_{b}| 0; t_{a} \rangle = \langle 0|
U(t_b,t_a) |0 \rangle \, ,
\end{displaymath}
\noindent and so for  time independent ${\hat{H}}$ one has
$F[t_{a}, t_{b}] = F[t_{b} -t_{a}]$.

\vspace{3mm}

\noindent The most usual way of calculating the fluctuation factor is
via the Van Vleck--Pauli--Morette determinant\cite{WP1,CDWM1,JVV1}:
\bea
&& F[t_{a},t_{b}] = \sqrt{{\det}_{2}\left(\frac{i}{2\pi \hbar}\,
\frac{\partial^{2} S_{cl}}{\partial {\bf x}^{\alpha}_{a} \partial {\bf
x}_{b}^{\beta} } \right)} \non \\ &&= \sqrt{{\det}_{2} \left(
\frac{i}{2\pi \hbar}\, \frac{\partial {\bf p}_{a\, \alpha}}{ \partial
{\bf x}_{b}^{\beta}} \right)}
 =\sqrt{{\det}_{2} \left(
\frac{i}{2\pi \hbar}\, \frac{\partial {\bf  p}_{b\, \alpha}}{ \partial
{\bf x}_{a}^{\beta}} \right)}\, .
\label{VPM}
\eea
\noindent The symbol $\det_{2}(\ldots)$ denotes $2\times2$ determinant.
In (\ref{VPM}) we have also used the identities ${\bf p}_{a} =
-\frac{\partial S_{cl}}{\partial {\bf x}_{a}}$ and ${\bf p}_{b} =
\frac{\partial S_{cl}}{\partial {\bf x}_{b}}$ (one should take a
little care when using the covariant and contravariant
indices). Indices $a$ and $b$ are kept fixed throughout calculation.

\vspace{3mm}

\noindent Actually (\ref{VPM}) is correct only for sufficiently short
elapsed times $t_b - t_a$ as it was proved by Pauli\cite{WP2}.  In the
general case, the determinant on the RHS of (\ref{VPM}) will become
infinite every time  the classical (position space) orbit touches
(or crosses) a caustic. A detailed examination of quadratic systems
reveals\cite{RGL1,HK1}  that (\ref{ker111}) remains valid even after
passing through the caustic, provided we write the fluctuation factor as
\begin{equation}
F[t_{a},t_{b}] = \sqrt{\left|\ {\det}_{2} \left(
\frac{i}{2\pi \hbar}\, \frac{\partial {\bf  p}_{b\, \alpha}}{ \partial
{\bf x}_{a}^{\beta}} \right)\right|}\, ,
\end{equation}
\noindent and insert a factor $\exp(-i\pi/2)$ for every reduction of the
rank of $1/\det_2\left(\partial^{2} S_{cl}/\partial {\bf x}^{\alpha}_{a}
\partial {\bf x}_{b}^{\beta} \right)$  at the caustic.
Thus we have ($t_{b}>t_{a}$):
\begin{eqnarray}
\langle {\bf x}_{b};t_{b}| {\bf x}_{a};t_{a} \rangle &=&  e^{-i
\frac{\pi}{2} n_{a,b}}\,F[t_{a},t_{b}] \, \mbox{exp}\left(
\frac{i}{\hbar}S_{cl}[{\bf x}] \right)
\, .
\label{ker112}
\end{eqnarray}
\noindent  Here $n_{a,b}$ is the Morse (or Maslov)
index\cite{VPM1,RGL1,HK1,MM1,RGL3,JM1}  of the classical path running
from ${\bf{x}}_a$ to ${\bf{x}}_b$ \footnote{The set of all points where
the inverse of the Van Vleck--Pauli--Morette determinant vanishes is
called a caustic. The Morse index then counts how many times the
classical orbit crosses (or touches) the caustic  when
passing from the initial to the final position. In the
literature, crossing  points are  often called focal or conjugate points.}.
The form
(\ref{ker112}) is due to Gutzwiller\cite{MCGu1} and the prescription
(\ref{ker112}) is nothing but the connection formula for relating the
kernels on both sides of the caustic in a continuous
way\cite{VPM1}. To simplify the discussion we  omit for a while
the delicate issue  of  caustics assuming that the determinant in (\ref{VPM})
is positive. We shall, however, return  to it
in Sections IV and V.

\vspace{3mm}

\noindent Now we are ready to calculate the fluctuation factor.
A little algebra gives us
\bea
&&F[t_{a}, t_{b}] = \frac{m}{4\pi \hbar D}\left[- \left(
\dot{B}^{1}_{1}(t_{b})\dot{B}^{1}_{2}(t_{b}) -
\dot{B}^{1}_{3}(t_{a})\dot{B}^{1}_{2}(t_{b}) \right.\right. \non \\
&&\mbox{\hspace{6mm}}\left.\left. +
\dot{B}^{1}_{1}(t_{b})\dot{B}^{1}_{3}(t_{a}) -
\dot{B}^{1}_{3}(t_{a})\dot{B}^{1}_{3}(t_{a}) +
\dot{B}^{2}_{4}(t_{a})\dot{B}^{2}_{1}(t_{b}) \right.\right. \non \\
&&\mbox{\hspace{6mm}} \left.\left.+
\dot{B}^{2}_{4}(t_{a})\dot{B}^{1}_{4}(t_{a}) -
\dot{B}^{2}_{2}(t_{b})\dot{B}^{2}_{1}(t_{b}) -
\dot{B}^{2}_{2}(t_{b})\dot{B}^{1}_{4}(t_{a})
\right)\right]^{\frac{1}{2}} \non  \\  &&\mbox{\hspace{12mm}} =
\frac{m}{2\pi \hbar} \sqrt{\frac{W}{D}} \, .
\label{ker3}
\eea
Here we have used the equations of the motion and the fact that
the Wronskian is time independent.
Since the kernel is uniquely determined from the classical
action, our argument on the uniqueness of the classical action implies
that $\langle {\bf x}_{b};t_{b}| {\bf x}_{a};  t_{a} \rangle$ does not
depend on the choice of a fundamental system. Note also that
due to the fact that $F[t_{a},t_{b}] = F[t_{b}-t_{a}]$, it follows from
(\ref{ker3}) that $D(t_{a},t_{b}) = D(t_{b}-t_{a})$.

\section{Wave functions for Bateman's  dual system}
\subsection{Wave functions $\psi_{n,l}(r,u,t)$ and $\psi_{n,l}(r,t)$}

\noi In order to calculate the wave function it is useful to
rewrite the kernel in hyperbolic polar coordinates $(r,u)$, with
$x_{1} = r\,\mbox{cosh} \, u$ and $x_{2} =
r\,\mbox{sinh}\,u$,
and then apply the defining relation\cite{RPF1}:
\begin{eqnarray}
\langle r_{b},u_{b};t_{b}| r_{a},u_{a};t_{a} \rangle &=& \sum_{n,l}
\psi_{n,l}(r_{b}, u_{b},t_{b})\psi^{(*)}_{n,l}(r_{a}, u_{a}, t_{a})\, ,
\non \\ t_{b} & > & t_{a}\, .
\label{wf1}
\end{eqnarray}
\noindent Here we have used the symbol $(*)$ instead of the usual
complex conjugation symbol $*$ - the need for this refinement will
show up in the following. Invoking (\ref{ker05}), (\ref{Scl1}) and
(\ref{ker3}) (see also Appendix C) we obtain:
\bea &&\langle r_{b}, u_{b}; t_{b}| r_{a}, u_{a}; t_{a} \rangle\non \\
&&\mbox{\hspace{0.5cm}}=\frac{m}{2\pi  \hbar} \sqrt{\frac{W}{D}}\,
\mbox{exp}\left[ \frac{i\, m}{2D \, \hbar} \left( -r_{a}^2 {\dot
B}_{1}^{1}(t_{a}) + r_{b}^{2} {\dot B}_{3}^{1}(t_{b})
\right. \right.\non \\  &&\mbox{\hspace{0.5cm}} \left.\left. + \,2
r_{a}r_{b}\,{\dot B}_{1}^{1}(t_{b})\mbox{cosh}(\Delta u) - 2
r_{a}r_{b}\,{\dot B}_{1}^{2}(t_{b})\, \, \mbox{sinh}(\Delta u)
\right)\right]\,  \non \\ [4mm]
&&\mbox{\hspace{0.3cm}} =\frac{m}{2\pi  \hbar} \sqrt{\frac{W}{D}}\,
\mbox{exp}\left[ \frac{i\, m}{2\, \hbar} \left( - \left[ r_{a}^{2}  +
r_{b}^{2}  \right] \frac{{\dot
B}^{1}_{1}(t_{a})}{D}\right. \right. \non \\ &&\mbox{\hspace{0.5cm}}
\left.\left. + \,2 r_{a}r_{b}\,\left[\frac{{\dot B}_{1}^{1}(t_{b})}{D}
\, \mbox{cosh}(\Delta u)  -  \frac{{\dot B}_{1}^{2}(t_{b})}{D}\, \,
\mbox{sinh}(\Delta u)\right] \right)\right]\, , \non \\  &&\non \\
&&\mbox{\hspace{2cm}} \Delta u = u_{b} - u_{a}; \;\;\; t_{b} > t_{a}\,
,
\label{ker1}
\eea
\noindent  By observing that
\bea [{\dot B}_{1}^{1}(t_{b})]^2 -[{\dot B}_{1}^{2}(t_{b})]^2 = W D\,
, \eea
\noindent we can put
\begin{eqnarray*}
\frac{{\dot B}_{1}^{1}(t_{b})}{D}&=&\sqrt{\frac{W}{D}} \cosh \alpha\,
, \\ \frac{{\dot B}_{1}^{2}(t_{b})}{D}&=&\sqrt{\frac{W}{D}} \sinh
\alpha\, .
\end{eqnarray*}
\noindent In Appendix D  we show that
\begin{equation}
\alpha(t_{a},t_{b}) =\Gamma\, (t_{a} - t_{b}) + \beta\, .
\label{alpha1}
\end{equation}
\noindent Here $\Gamma = \frac{\gamma}{2m}$ and $\beta$ is a {\em
complex} constant. It is also useful to denote the
reduced oscillators frequency as $\Omega =
\sqrt{\frac{1}{m}\left(\kappa -\frac{\gamma^{2}}{4m}\right)}$. If not
indicated otherwise, $\Omega$ will be assumed to be real
throughout. That is, we shall mostly be concerned with the
under--damped case although occasionally a result can be taken over to
the over--damped case.

\vspace{3mm}

\noindent Eq.(\ref{alpha1}) allows to rewrite the kernel (\ref{ker1})
in the following form
\begin{eqnarray}
&&\langle r_{b}, u_{b}; t_{b}| r_{a}, u_{a}; t_{a} \rangle  \nonumber
\\  &&\mbox{\hspace{5mm}}=\frac{m}{2\pi \,  \hbar}
\sqrt{\frac{W}{D}}\, \mbox{exp}\left[ - \frac{i \, m}{4D \, \hbar} \,
\left( \frac{d D}{d t_{a}} \, r_{a}^2  - \frac{d D}{d t_{b}}\,
r_{b}^{2}\right) \ri]  \nonumber\\   &&\mbox{\hspace{1cm}}  \times
\mbox{exp}\left[\frac{i\, m}{ \hbar}
\,\sqrt{\frac{W}{D}}\,r_{a}r_{b}\,\cosh(\Delta u - \alpha) \right]\, .
\label{radker1}
\end{eqnarray}
\noindent This expression  may be recast into a more suitable
form if we apply the Laurent expansion\cite{GR1,MF1}\footnote{Because
$\mbox{exp}(i\,a \, \mbox{cosh}(u))$ is an analytic function
of $u$ - the only essential
singularities are in $u= \pm \infty$ - the Laurent
expansion (\ref{bes12}) is well defined for any complex $u$.}
\begin{equation} \mbox{exp}(i\,a \, \mbox{cosh}(u))  =   \sum_{l=
-\infty}^{\infty} (-1)^{l}\, I_{l}(-i\,a)\, e^{-lu} \, ,
\label{bes12}
\end{equation}
\noindent ($I_{l}(\ldots)$ are the modified (or hyperbolic) Bessel
functions) together with the addition theorem for the generalized
Laguerre polynomials $L^l_n$ (Mehler's formula\cite{GR1,MF1}):
\begin{eqnarray}
&&\sum_{n=0}^{\infty} n! \frac{L^l_n(z_1) L^l_n(z_2) b^n}{\Ga
(n+l+1)}\nonumber \\ && \mbox{\hspace{5mm}} = \,\frac{(z_1 z_2
b)^{-\frac{1}{2}l}}{1 -b}\exp\lf[-b\,\frac{z_1+ z_2}{1-b}\right]
\, I_l \lf(2\frac{\sqrt{z_1 z_2 b}}{1-b}\ri)\, . \label{laguerre}
\end{eqnarray}
\noindent For this purpose we set
\begin{displaymath}
z_1 = \frac{m}{\hbar}\, \sqrt{W}\, \frac{r_a^2}{\rho(t_{a})}\, ,
\;\;\;\; z_2 = \frac{m}{\hbar}\, \sqrt{W}\,
\frac{r_b^2}{\rho(t_{b})}\, ,
\end{displaymath}
\noindent with $r_{a}^{2}; r^{2}_{b} \geq 0$\footnote{Because
$\cosh(u)\geq \sinh(u)$ then $|x_1| \geq |x_2|$ and so we
have automatically that $r^2 \geq 0$ is a
kinematic invariant.} and
\begin{eqnarray}
\rho(t) &=& V(t) W \left( \int \frac{dt}{V(t)} \right)^{2} +
V(t)\nonumber \\ &=& \sqrt{\sum_{i<j}^{4}\, \left({\bf u}_{i}(t)
\wedge {\bf u}_{j}(t)\right)^{2}  } \, .
\label{rho1}
\end{eqnarray}
\noindent This allows to identify the parameter $b$ appearing in
(\ref{laguerre}) with $b(t_{b})$ where $b(t)$ reads
\begin{eqnarray*}
b(t) &=& \frac{-i \sqrt{V(t)}   + \sqrt{\rho(t) - V(t)}}{i \sqrt{V(t)}
+ \sqrt{\rho(t) - V(t)}}\\  &=& \frac{\left( -i\sqrt{V(t)} +
\sqrt{\rho(t) - V(t)}\; \right)^{2}}{\rho(t)}\\ &=&
\mbox{exp}\left(-i2 \;
\mbox{arcsin}\sqrt{\frac{V(t)}{\rho(t)}}\;\right)\, .
\end{eqnarray*}
\noindent where $V(t) = {\bf u}_{3}(t)\wedge {\bf u}_{4}(t) =  {\bf
v}_{1}(t)\wedge {\bf v}_{2}(t)$.
Note that $\rho(t_{a}) = U_{a}$  and that $b(t_{a}) =1$.
\vspace{3mm}

\noindent The previous manipulations permit us to formulate the kernel  in
the desired form
\begin{eqnarray}
&&\langle r_{b}, u_{b};t_{b}|r_{a}, u_{a};t_{a} \rangle
= \frac{i}{\pi} \sum_{n,l}\frac{n!}{\Ga
(n+l+1)}
\nonumber \\
&&\mbox{\hspace{3mm}}  \times \,
\lf[\,b^{*}(t_{a})b(t_{b} \ri)\,]^{n+\frac{l+1}{2}} \lf(\frac{m}{
\hbar}\, \sqrt{\frac{W}{\rho(t_{a})\rho(t_{b})}}\ri)^{l+1}
\nonumber \\
&&\mbox{\hspace{3mm}} \times
L^l_n\left(\frac{m}{\hbar}\sqrt{W}\,\frac{r_a^2}{\rho(t_{a})}\right)\,
L^l_n\left(\frac{m}{\hbar}\sqrt{W}\,\frac{r_b^2}{\rho(t_{b})}\right)
(r_a r_b)^{l}
\nonumber \\  &&\mbox{\hspace{3mm}} \times \,
\mbox{exp}\left(\frac{m}{2\hbar}\left[ \frac{i}{2} \frac{\dot{\rho}
(t_{b})}{\rho(t_{b})} - \frac{\sqrt{W}}{\rho(t_{b})}
\right]r^{2}_{b}\right. \nonumber \\ &&\mbox{\hspace{11mm}} -\,
\left. \frac{m}{2\hbar}\left[ \frac{i}{2}
\frac{\dot{\rho}(t_{a})}{\rho(t_{a})} + \frac{\sqrt{W}}{\rho(t_{a})}
\right]r^{2}_{a}   \right) \, e^{l(u_a -u_b + \alpha(t_{a},t_{b}))}\,
.  \nonumber \\
\label{ker11}
\end{eqnarray}
\noindent Let us identify the wave function $\psi_{n,l}(r, u, t) =
\langle r, u| \psi_{n,l}(t) \rangle$. Note that Eq.(\ref{ker11})
immediately implies that $\psi^{(*)}_{n,l}(r,u,t)$ cannot be
associated with $\psi^{*}_{n,l}(r,u,t)$. It is not difficult to see
that this peculiar behavior goes into account of the seemingly
harmless expansion (\ref{bes12}). The point is that we have tacitly
used  the discrete (Laurent) expansion even if an alternative integral
(continuous) expansion was available\cite{GR1}. This favoritism towards
discrete $l$'s was
deliberate (see also next Section).
A careful analysis will reveal
that the discreteness of $l$ is not
compatible with a unitary representation of the
dynamic symmetry group of the theory. The remedy will be found in a
self--adjoint extension of ${\hat{H}}$ and it
will turn out that $\psi^{(*)}_{n,l}(r,u,t) = \psi_{n,l}(r, -u,
-t)$.

\vspace{3mm}

\noindent Now, from (\ref{wf1}) and (\ref{ker11}) we may deduce the wave
functions
\begin{eqnarray}
&&\psi_{n,l}(r,u,t) = \sqrt{\frac{1}{\pi}}
\sqrt{\frac{n!}{\Gamma(n+l+1)}} \left(\sqrt{\frac{m}{\hbar \,\rho(t)}}
\, W^{1/4} \right)^{l+1}\nonumber \\ &&\mbox{\hspace{5mm}}\times \;
[\,b(t)\,]^{n + \frac{l+1}{2}}\, L^{l}_{n}\left(
\frac{m}{\hbar} \sqrt{W}\, \frac{r^{2}}{\rho(t)} \right) \,r^{l}\nonumber \\
&&\mbox{\hspace{5mm}}\times \;  \mbox{exp}\left(\frac{m}{2\hbar}
\left[\frac{i}{2}  \frac{\dot{\rho}(t)}{\rho(t)} -
\frac{\sqrt{W}}{\rho(t)}\right] r^{2} \right) e^{-l(u + \Gamma t -
\frac{\beta}{2})}\, ,\nonumber
\eea
\bea
&&\psi^{(*)}_{n,l}(r,u,t) = \sqrt{\frac{1}{\pi}}
\sqrt{\frac{n!}{\Gamma(n+l+1)}} \left(\sqrt{\frac{m}{\hbar \,\rho(t)}}
\, W^{1/4} \right)^{l+1}\nonumber \\ &&\mbox{\hspace{5mm}}\times \;
[\,b^{*}(t)\,]^{n + \frac{l+1}{2}}\,  L^{l}_{n}\left(
\frac{m}{\hbar} \sqrt{W}\, \frac{r^{2}}{\rho(t)} \right)\,r^{l} \nonumber \\
&&\mbox{\hspace{5mm}}\times \;  \mbox{exp}\left(-\frac{m}{2\hbar}
\left[\frac{i}{2}  \frac{\dot{\rho}(t)}{\rho(t)} +
\frac{\sqrt{W}}{\rho(t)}\right] r^{2} \right) e^{l(u + \Gamma t -
\frac{\beta}{2})} \, ,\nonumber \\
\label{wave}
\end{eqnarray}
%
%
%
%


\noindent Obviously, neither $\psi_{n,l}(r,u,t)$ nor
$\psi^{(*)}_{n,l}(r,u,t)$ belong to ordinary Hilbert space because
they cannot be normalized in the usual manner (they do not belong
to the space of square  integrable functions $\ell^2$). The latter
observation is in agreement with Refs.\cite{HD1,GV1}, and we shall
comment more on this point in the next subsection. We note that
the kernel (\ref{ker1}) (and consequently the wave functions
(\ref{wave})) satisfies the time--dependent Schr{\"o}dinger
equation\footnote{Actually $\psi^{(*)}_{n,l}(r,u,t)$ fulfills the
time--reversed (time--dependent) Schr{\"o}dinger
equation, see Eq.(\ref{timerevsch}).}:
\begin{displaymath}
\left( i\hbar\, \frac{\partial }{\partial t_{b}} - {\hat{H}}(r_{b},
u_{b})\right) \langle r_{b}, u_{b}; t_{b}| r_{a}, u_{a}; t_{a} \rangle
= 0\, , \, \;\;\; t_{b} > t_{a} \, ,
\end{displaymath}
\noindent where
\begin{eqnarray} \label{Ham1}
&&{\hat{H}} = \frac{1}{2 m} \left[{\hat{p}}_{r}^{2} -
\frac{1}{r^{2}}{\hat{p}}_{u}^{2} + m^2 \Omega^2 r^{2} \right] - \Gamma
{\hat{p}}_{u} \\ && \non
\\ \non &&= \frac{1}{2 m} \left[- \hbar^2
\frac{\partial^{2} }{\partial r^{2}} -  \frac{\hbar^2}{r}
\frac{\partial}{\partial r} +  \frac{\hbar^2}{r^2} \frac{\partial^{2}
}{ \partial u^2}
+ m^2 \Omega^2 r^2 \right]
+ i \hbar \Gamma \frac{\partial }{\partial u}\, .
\end{eqnarray}
\noindent The Hamiltonian (\ref{Ham1}) is the so called Bateman
Hamiltonian\cite{HD1}. We now define the {\em radial} kernel $\langle
r_{b};t_{b}|r_{a};t_{a} \rangle_{n,l}$ as\cite{HK1}
\begin{displaymath}
\langle r_{b},u_{b};t_{b}| r_{a}, u_{a}; t_{a} \rangle = \sum
_{n,l}\frac{ \langle r_{b};t_{b}| r_{a};t_{a} \rangle_{n,l} }{\pi
\sqrt{r_{a} r_{b}}}\; e^{l(\alpha(t) - \Delta u)} \, .
\end{displaymath}
\noindent The corresponding wave function $\psi_{n,l}(r,t) = \langle r
| \psi_{n,l}(t) \rangle$ reads
\begin{eqnarray}
&&\psi_{n,l}(r,t) = \sqrt{\frac{n!}{\Gamma(n+l +1)}}
\left(\sqrt{\frac{m}{\hbar \, \rho(t)}}\, W^{1/4}
\right)^{l+1}\nonumber \\ &&\mbox{\hspace{5mm}}\times \; [\,b(t)\,]^{n
+ \frac{l+1}{2}}\,r^{l +  \frac{1}{2}}\, L^{l}_{n}\left(
\frac{m}{\hbar} \sqrt{W}\, r^{2}/\rho(t) \right) \nonumber \\
&&\mbox{\hspace{5mm}}\times \;  \mbox{exp}\left(\frac{m}{2\hbar}
\left[\frac{i}{2}  \frac{\dot{\rho}(t)}{\rho(t)} -
\frac{\sqrt{W}}{\rho(t)}\right] r^{2} \right)\, .
\label{wave2}
\end{eqnarray}
\noindent It is simple to persuade oneself that
$\psi_{n,l}(r,t)$ fulfils  both the orthonormalization condition
\begin{displaymath}
\int_{-\infty}^{\infty} dr\, \psi^{(*)}_{n,l}(r,t)\,\psi_{n',l}(r,t) =
\delta_{nn'}\, ,
\end{displaymath}
\noindent and the resolution of unity
\begin{displaymath}
\sum_{n=0}^{\infty} \psi^{(*)}_{n,l}(r,t)\,\psi_{n,l}(r',t) =
\delta(r-r')\, .
\end{displaymath}
\noindent Note that both the radial kernel and the wave function
(\ref{wave2}) satisfy the time--dependent Schr{\"o}dinger equation
\begin{displaymath}
\left( i\hbar \,\frac{\partial}{\partial t_{b}} - {\hat H}_{l}(r_{b})
\right)  \langle r_{b};t_{b}|r_{a};t_{a} \rangle_{n,l} = 0\, ,
\;\;\;\; t_{b} > t_{a}\, ,
\end{displaymath}
\noindent where
\begin{eqnarray}
{\hat H}_{l} &=& \frac{1}{2m} \left[ - \hbar^2  \frac{\partial^2
}{\partial r^{2}} + \frac{\hbar^2}{r^2} \, \left( l^{2} - \frac{1}{4}
\right) + m^{2} \Omega^2 r^{2}\right]\nonumber \\ &-& i\hbar {\dot
\alpha(t)}\, l  - i\hbar \Gamma  \, l \, .
\label{Hamilt2}
\end{eqnarray}
%
%
\noindent The term proportional to $1/r^2$ is analogous to the {\em
centrifugal barrier} known from rotationally invariant systems and so
the quantum number $l$ can be viewed as analog of the {\em
azimuthal} quantum number. Note that, due to the structure of
$\alpha(t_{a},t_{b})$, the term ${\dot \alpha}(t) + \Gamma$ must be
zero.

\vspace{3mm}

\noindent Since the generalized
Laguerre polynomials $L_{n}^{l}$ are defined
for all $l\in \mathbb{C}$ indices\footnote{Analytic continuation
for $l$ with $\Re(l) = -1,-2,-3, \ldots$ is however required, see
e.g.\cite{Fl1}}, the wave functions (\ref{wave2}) satisfy the
time--dependent Schr{\"o}dinger equation with the Hamiltonian
(\ref{Hamilt2}) also for non--integer $l$'s. The key observation
then is that if we continue $l$ to the values $\pm \frac{1}{2}$,
the Hamiltonian ${\hat H}_{l}$ describes the 1D l.h.o..
If we make use of the rules connecting Hermite
polynomials with  the $l=\pm \frac{1}{2}$ Laguerre
polynomials\cite{GR1}, we may rewrite the continued radial wave
functions in a simple form
\begin{eqnarray}
&&\psi_{n,\frac{1}{2}}(r,t) = \frac{1}{2^{2n+1}}\sqrt{\frac{1}{n!\,
\Gamma\left(n + \frac{3}{2} \right)}} \left(\sqrt{\frac{m}{\hbar
\,\rho(t)}} \, W^{\frac{1}{4}} \right)^{\frac{1}{2}}\nonumber \\
&&\mbox{\hspace{14mm}}\times \; [\,b(t)\,]^{n + \frac{3}{4}} \,
H_{2n+1}\left( \sqrt{\frac{m}{\hbar \,\rho(t)}}\,  W^{\frac{1}{4}}\, r
\right) \nonumber \\ &&\mbox{\hspace{14mm}}\times \;
\mbox{exp}\left(\frac{m}{2\hbar} \left[\frac{i}{2}\,
\frac{\dot{\rho}(t)}{\rho(t)} - \frac{\sqrt{W}}{\rho(t)}\right] r^{2}
\right)\, ,\nonumber
\eea
\bea
&&\psi_{n,-\frac{1}{2}}(r,t) = \frac{1}{2^{2n}}\sqrt{\frac{1}{n!\,
\Gamma\left( n + \frac{1}{2} \right)}} \left(\sqrt{\frac{m}{\hbar\,
\rho(t)}}\, W^{\frac{1}{4}}  \right)^{\frac{1}{2}}\nonumber \\
&&\mbox{\hspace{16mm}}\times \; [\,b(t)\,]^{n + \frac{1}{4}} \,
H_{2n}\left(\sqrt{ \frac{m}{\hbar \, \rho(t)}} \, W^{\frac{1}{4}}\, r
\right) \nonumber \\ &&\mbox{\hspace{16mm}}\times \;
\mbox{exp}\left(\frac{m}{2\hbar} \left[\frac{i}{2}\,
\frac{\dot{\rho}(t)}{\rho(t)} - \frac{\sqrt{W}}{\rho(t)}\right] r^{2}
\right)\, .
\label{wave4}
\end{eqnarray}
\noindent In passing we mention that the quantum numbers $n$ and $l$
appearing in (\ref{ker11})--(\ref{wave}) have been seemingly
independent. So far the only obvious restriction was that $n \geq 0$
integers.
However, for a consistent probabilistic interpretation
(in the $r$ variable) and analytical
continuation (\ref{wave4}), the wave function
$\psi_{n,l}(r,u,t)$ is required to be bounded for $|r| < \infty$.
Using the asymptotic expansion for $L^l_n(z)$ (see
e.g.\cite{GR1,MF1}):
\begin{eqnarray*}
L^l_n(z) &=& \frac{[\Gamma(n+l+1)]^2}{n!\, \Gamma(l+1)}\,
{}_1F_1(-n,l+1;z)\nonumber \\  &\approx & \frac{\Gamma(n+l+1)}{n!}(-z)^n;
\;\;\;\;\; z \rightarrow \infty \nonumber \\ &\approx & 1 -\frac{n}{l}
\, z; \;\;\;\;\; z \rightarrow 0 \, ,
\end{eqnarray*}
\noindent where ${}_1F_1(-n,l+1;z)$ is the confluent hypergeometric
series\cite{MF1}. It is not difficult to see that
the only allowed values of $n$ and $l$  at which
$\psi_{n,l}(r,u,t)$ fulfills above requirements are those where:
\begin{equation}
2n + l + 1 = 0, \pm 1, \pm 2, \,  \dots\, , \;\;\;\; l \not= 0 \, .
\label{disc1}
\end{equation}
\noindent So $|l|  \geq 1$ and $n \geq 0$.

\subsection{Meaning of quantum numbers $n$ and $l$}

\noindent Let us now consider the meaning of the quantum numbers $n$
and $l$. To do this we must first understand the algebraic structure
of the Hamiltonian (\ref{Ham1}). In Refs.\cite{HD1,GV1} the following
ladder operators were introduced:
\bea A&=&\frac{1}{\sqrt{2\hbar m\Om}}\lf[{\hat p}_1 - i m \Om x_1
\ri]\, ,\nonumber \\   B&=&\frac{1}{\sqrt{2\hbar m\Om}}\lf[{\hat p}_2
- i m \Om x_2 \ri]\, ,  \lab{aboperators}
\label{nl2}
\eea
\noindent with
\begin{equation}
\left[A, A^{\dagger}\right] = \left[B, B^{\dagger} \right] = 1, \;\;\;
\left[A, B \right] = \left[A, B^{\dagger} \right] = 0 \, .
\label{n121}
\end{equation}
\noindent The Hamiltonian (\ref{Ham1}) can be then rewritten as
\bea {\hat H}&=& \hbar \Omega(A^{\dagger}A - B^{\dagger}B) + i\hbar
\Gamma(A^{\dagger}B^{\dagger} -AB )\nonumber \\ &=& 2\hbar (\Omega
{\cal{C}} - \Gamma J_{2})\, ,
\label{nl3}
\eea
\noindent where we have made explicit the associated
$SO(2,1)\equiv SU(1,1)$
algebraic structure:
\bea {\cal{C}}^{2} &=& \frac{1}{4}(A^{\dagger}A - B^{\dagger}B)^{2}\,
,\nonumber \\  J_{+} &=& A^{\dagger}B^{\dagger} \, , \;\;\;\; J_{-} =
A B\, , \nonumber \\  J_{3} &=& \frac{1}{2}(A^{\dagger}A +
B^{\dagger}B + 1)\, ,\nonumber \\  &&\mbox{\hspace{-1.6cm}} [ J_{+},
J_{-}] = - 2J_{3}\, , \;\;\;\; [J_{3},J_{\pm}]  = \pm J_{\pm} \, .
\label{nl6}
\eea
\noindent Here ${\cal{C}}$ is the only Casimir operator ( $SU(1,1)$
 has rank 1). In addition, $[{\cal{C}},{\hat H}] =  [J_{2},
{\hat H}] = 0$. If one defines
\begin{eqnarray}
J_{1}  = \frac{1}{2}(J_+ + J_-)\, \quad,\quad
J_{2}  =
-\frac{i}{2}(J_+ - J_-)\, ,
\label{n161}
\end{eqnarray}
\noindent the more familiar $SU(1,1)$ algebraic structure appears:
\begin{displaymath}
[ J_{1},J_{2}] = -iJ_{3}\, ,\;\;\; [ J_{3},J_{2}] = -iJ_{1}\, , \;\;\;
[ J_{1},J_{3}] = -iJ_{2}\, ,
\end{displaymath}
\noindent with
\begin{equation}
{\cal{C}}^2 = J_3^2 - J^2_2 - J_1^2 + \mbox{$\frac{1}{4}$}\, .
\end{equation}
\noindent It is simple to check that
\bea\non
{\cal{C}} &=& \frac{1}{4\hbar \Omega m}\left[{\hat{p}}_{r}^{2} -
\frac{1}{r^{2}}{\hat{p}}_{u}^{2} + m^2 \Omega^2 r^{2}  \right] \,,
\\ \non
 J_{2} &=& \frac{1}{2\hbar}\, {\hat p}_{u} \, .
\eea
\noindent In this connection it is important to recognize that the
system described by the Hamiltonian (\ref{nl3}) is both conservative
and invariant under time reversal. Because the latter point has been
treated in the literature in a somehow ambiguous fashion
(cf. Ref.\cite{HD1}), we discuss it in detail in Appendix E. We prove there
that ${\cal{T}} {\cal{C}}{\cal{T}}^{-1} = {\cal{C}}$ and  ${\cal{T}}
J_{2} {\cal{T}}^{-1} =  J_{2}$ from which it follows that
${\cal{T}}{\hat H}{\cal{T}}^{-1}= {\hat H}$.   It is
precisely this time--reversal issue which obscures the  quantization of
Bateman's  system, bringing about many subtleties which are difficult
to grasp without an explicit  knowledge of the time--dependent
wave functions (\ref{wave}).

\vspace{3mm}

\noindent Now, the crucial observation is that although from (\ref{n161}),
i.e. from the very definition, $J_{2}$ appears to be  Hermitian,
(\ref{wave}) implies that it has a purely imaginary
spectrum in $|\psi_{n,l}(t) \rangle$ (and this holds for all $t$). The
root of this ``pathological'' behavior is in the non--existence of a
unitary irreducible representation of $SU(1,1)$ in which $J_{2}$ would
have at the same time a real and discrete
spectrum\cite{GL1}. Nevertheless, both the discreteness and
complexness of $J_2$ spectra are vital in our analysis since they bring
dissipative features in the dynamics: this was also the
case considered in Refs.\cite{HD1,HF1,GV1}, to which we are going to compare
our results.
Thus  the usual unitary representations of $SU(1,1)$
are clearly not useful for our purpose. On the other hand,
by resorting to
non--unitary representations of $SU(1,1)$  (known as  non--unitary
principal series\cite{APE1,AWK1}) we lose the
hermiticity\footnote{To be precise, we should talk about
self--adjointness rather than hermiticity, but we shall assume here
and throughout that this ambiguity
does not cause any harm in the  present context.} of $J_2$ and hence
the spectral theorem along with the resolution of unity.

\vspace{3mm}

\noindent  Actually the situation is not so hopeless.  One may indeed
redefine the inner product\cite{GV1,AWK1} to get  a unitary
irreducible representation (known as complementary
series\cite{APE1,AWK1}) out of non--unitary principal series.  This
may be easily done when we notice, using (\ref{wf1}), (\ref{wave}) and
Schwinger's prescription (\ref{ker03}), that the
states $\psi^{(*)}_{n,l}$
are not simple complex conjugates of $\psi_{n,l}$ because they fulfil
the time--dependent Schr{\"o}dinger equation
\begin{equation}\lab{timerevsch}
\left( i\hbar \, \frac{\partial}{\partial t} +
{\cal{T}}{\hat{H}}^{\dagger}(r,u)
{\cal{T}}^{-1}\right)\, \psi^{(*)}_{n,l}(r,u,t) = 0\, ,
\end{equation}
\noindent with the (effectively) non--Hermitian Hamiltonian.
Accordingly, what we have loosely denoted in
(\ref{wf1}) as $\psi_{n,l}^{(*)}(r,u,t)$ is
actually
\begin{equation}
\langle{\cal{T}}\, \psi_{n,l}(-t) |r,u \rangle = \langle
\psi_{n,l}(-t)| r,-u \rangle^{*} = \psi_{n,l}(r, -u, -t)\, ,
\label{wave12}
\end{equation}
\noindent as can be also double--checked from the explicit form
(\ref{wave}). For the sake of simplicity we use
$[{\cal{T}}\,|\psi_{n,l}(t) \rangle]^{\dagger} =\langle{\cal{T}}\,
\psi_{n,l}(t)| $. Clearly, if $J_{2}$ were Hermitian
then  $\psi^{(*)}_{n,l} = \psi^{*}_{n,l}$ as one would expect.

\vspace{3mm}

\noindent The above considerations have some important implications.
To see this, let us rewrite the kernel (\ref{ker11})
by means of the states $|\psi_{n,l}(t)\rangle$:
\begin{eqnarray}
&&\langle r_{b},u_{b};t_{b}|r_{a},u_{a};t_{a}\rangle
= \sum_{n,l}\psi_{n,l}(r_b,u_b,t_b)
\,\psi_{n,l}^{(*)}(r_a,u_a,t_a)  \nonumber \\ &&\mbox{\hspace{7mm}}=
\sum_{n,l} \langle r_{b},u_{b}|\psi_{n,l}(t_b)\rangle \langle
{\cal{T}}\, \psi_{n,l}(-t_{a})|r_{a},u_{a}\rangle \, .
\end{eqnarray}
\noindent We can  formally introduce the conjugation operation (``bra
vector'') as $\langle \psi_{n,l}(t)| \equiv [{\cal{T}}\,|\psi_{n,l}(-t)
\rangle]^{\dagger}$. Then the resolution of unity can be  written in a
deceptively simple form
\begin{equation}
\sum_{n,l}|\psi_{n,l}(t)\rangle \langle \psi_{n,l}(t)| = 1\, .
\label{h1}
\end{equation}
\noindent The price which has been paid for this simplicity  is that
we have endowed the Hilbert space  with a  new inner product.  In this
context two points should be stressed. First,  under the new inner
product $|\psi_{n,l}(t)\rangle$ has a finite (and positive)
norm. Second,  $J_2$ is Hermitian with respect
to this inner product. Indeed, integrating by parts we get
\begin{eqnarray}
&&\langle \psi_{n,l}(t)| J_{2}\,\psi_{m,k}(t) \rangle \nonumber \\
&&\mbox{\hspace{5mm}}= \, \int dr du \ r \, \psi_{n,l}^{(*)}(r,u,t)
\,\left( - \frac{i}{2} \frac{\partial }{\partial u} \right)
\psi_{m,k}(r,u,t)\nonumber \\ &&\mbox{\hspace{5mm}}= \,  \int dr du \
r  \left[\left(-\frac{i}{2} \frac{\partial }{\partial u}\right)\,
\psi_{n,l}(r,u,t)\right]^{(*)} \,   \psi_{m,k}(r,u,t)\nonumber \\
&&\nonumber \\ &&\mbox{\hspace{5mm}}=\langle J_{2}\,
\psi_{n,l}(t)|\psi_{m,k}(t)  \rangle \, .
\label{h11}
\end{eqnarray}
\noindent In (\ref{h11}) we have applied (\ref{TI1}), (\ref{wave12})
together with the fact  that the ``surface'' term is zero (if $k \not=
l$ then integration w.r.t. the variable $r$ gives zero\cite{GR1}, if $k=l$
then the product $\psi^{(*)}_{n,l}\,\psi_{m,l}$ is $u$ independent).

\vspace{3mm}

\noindent Note that the (\ref{h11}) implies that
$[{\cal{T}}J_2{\cal{T}}^{-1}]^{\dagger}=J_2$, or equivalently:
${\cal{T}}J_2^{\dag}{\cal{T}}^{-1}=J_2$.
Indeed,
\begin{equation}
\langle J_2 \psi_{n,l} (t)| \,=\, [{\cal{T}} J_2
|\psi_{n,l}(-t)\ran]^{\dagger}
\,= \, \langle\psi_{n,l} (t)| [{\cal{T}}J_2{\cal{T}}^{-1}]^{\dagger}\, .
\label{TII11}
\end{equation}
\noindent
The spurious time irreversibility of $J_2$  apparent in (\ref{TII11})
is an obvious consequence of dealing with the non--unitary
representation of $SU(1,1)$. This is understandable, since  a
mechanism  which does not preserve the norm (dissipation) is inherently
connected with time irreversibility. From a mathematical point of
view,  we can interpret the relation
${\cal{T}}J_2{\cal{T}}^{-1}=J^{\dagger}_2$ as a
self--adjoint extension of $J_2$ in the space spanned by
the $|\psi_{n,l}(t)\rangle$ vectors. It should be also clear that
when $J_2$ is time reversible (e.g., in the usual Hilbert space
$\ell^2$)  it is also automatically Hermitian.

\vspace{3mm}

\noindent A pivotal consequence of the above is that
\begin{eqnarray}
\langle \psi_{n,l}(t)| &=& \left[ {\cal{T}} |\psi_{n,l}(-t) \rangle
\right]^{\dagger} =
\left[{\cal{T}} e^{\frac{it}{\hbar}\, {\hat{H}}}\,|\psi_{n,l}(0) \rangle
\right]^{\dagger}\nonumber \\
 &=& \langle \psi_{n,l}(0)|\, e^{\frac{it}{\hbar} \,
{\hat{H}}}\, .
\end{eqnarray}
\noindent Thus the time--evolution operator is unitary under the
new inner product.
It is this unitarity condition, intrinsically
built in the kernel formula (\ref{ker06}) (and successively taken
over by the Feynman--Hibbs prescription), which naturally leads to a
``consistent'' inner product introduced in a somehow intuitive manner in
Refs.\cite{HD1,GV1}. From now on the modified inner product will be always
tacitly assumed.

%

\vspace{3mm}

\noindent So far we have dealt with the peculiar structure of the
Hilbert space. To interpret the quantum numbers $n,l$  labelling the
constituent states, we start with the  observation that from  the
explicit form (\ref{wave}) one can readily construct the Hermitian
operator  $\tilde{\cal{C}}$  (commuting with $J_2$) which is
diagonalized by $\psi_{n,l}(r,u,t)$.  Indeed one may check that
\begin{eqnarray}
&&J_{2}\,\psi_{n,l}(r,u,t) = -\frac{i}{2}\frac{\partial}{\partial u}\,
\psi_{n,l}(r,u,t) = i\,\frac{l}{2} \psi_{n,l}(r,u,t)\, ,\nonumber \\
&&{\tilde{\cal{C}}}\,\psi_{n,l}(r,u,t)=
\frac{\sqrt{W}}{2\Omega\rho}(2n + l + 1)  \;\psi_{n,l}(r,u,t)\, .
\label{qn1}
\end{eqnarray}
\noindent Here
\begin{eqnarray}
{\tilde{\cal{C}}} &=& {\cal{C}} - \frac{m}{4\Omega \hbar}\left(
\Omega^2 - \frac{W}{\rho^2} -\frac{{\dot{\rho}}^2}{4 \rho^2}\right)r^2
+  \,\frac{i}{4\Omega}  \frac{\dot{\rho}}{\rho}\, \left( r
\frac{\partial}{\partial r} +  1\right)\nonumber \\ &&\nonumber \\ &=&
e^{2\zeta}\; \hat{R}(t) \,{\cal{C}}\; \hat{R}^{-1}(t)\, .
\label{C1}
\end{eqnarray}
\noindent The unitary operator $\hat{R}(t)$ has the form
\begin{eqnarray*}
&&\hat{R}(t) =  \hat{S}(\xi;t)\, \mbox{exp}(i\zeta\, G_A)\,\mbox{exp}
(i\zeta\, G_B) \, ,\\ &&\hat{S}(\xi;t) =  \mbox{exp}\left( i \, \xi \,
r^2 \right)\, ,
\end{eqnarray*}
\noindent with
\begin{displaymath}
G_A = i\frac{1}{2}\left( A^2 - (A^{\dagger})^2 \right)\, , \;\; G_B =
i\frac{1}{2}\left( B^2 - (B^{\dagger})^2 \right)\, ,
\end{displaymath}
\noindent and
\begin{displaymath}
\zeta = \frac{1}{4}\, \mbox{ln}\left(\frac{W}{\Omega^2\rho^2} \right)
\, , \;\; \xi = \frac{m}{4 \hbar} \,  \frac{\dot{\rho}}{\rho}
 \;\;\; \Rightarrow \;\;\;\xi = - \frac{m}{2\hbar}\, \dot{\zeta} \, .
\end{displaymath}
\noindent The reader may recognize in $G_{A}$ and $G_{B}$  the
$SU(1,1)$ displacement operators (i.e., generalized coherent states
generators)\cite{APE1}. Alternatively, one may view $G_{A}$ and
$G_{B}$ as the single--mode squeeze operators\cite{MW1}.
One may also notice that $J_2$ is nothing but the
generator of two mode--squeeze\cite{MW1}. Actually, the fact that
there should be a close connection between $SU(1,1)$ squeezed states
and damped oscillators was firstly proposed in Ref.\cite{CRTV1}.
Finally we should point out that in deriving (\ref{C1}) the relation
$\hat{p}_r = -i\hbar\,\left(\partial/\partial r + 1/2r \right) $ was
used\cite{HK1}.

\vspace{3mm}

\noindent From Eq.(\ref{C1}) we find that $\tilde{\cal{C}}$ is
Hermitian whenever  $\rho$ is a real function and
${\cal{T}}\tilde{\cal{C}}(t){\cal{T}}^{-1} = \tilde{\cal{C}}(-t)$.
The latter together with (\ref{qn1})  implies that
\begin{equation}
{\cal{T}}\, |\psi_{n,l}(t)\rangle = |\psi_{n+l, -l}(-t) \rangle \, ,
\label{TR23}
\end{equation}
\noindent and so  in the static case (i.e. when $\rho(t) =
const.$ and $V(t)= \rho \; \mbox{sin}^{2}(\Omega t) $, see also
Section V) we have
\begin{eqnarray}
J_{2}\,|\psi_{n,l}^s\rangle &=& i\,\frac{l}{2}\,  |\psi_{n,l}^s\rangle
\, , \nonumber \\ {\cal{C}}\,|\psi_{n,l}^s \rangle &=& \frac{1}{2}
\,\left( 2n + l+1\right)  |\psi_{n,l}^s \rangle \, .
\label{state0}
\end{eqnarray}
\noindent Here (and throughout) the convention $|\psi_{n,l}^s \rangle
\equiv |\psi_{n,l}^s (0)\rangle $ is employed.
Notice that in view of relations (\ref{TI1}) and (\ref{TR23}),  the
time reversed stationary states fulfill the condition:
\begin{equation}
{\cal{T}}\,|\psi_{n,l}^s\rangle = |\psi_{n+l,-l}^s \rangle \, .
\label{TI11}
\end{equation}

\vspace{3mm}

\noindent Using (\ref{qn1}), (\ref{C1}) and (\ref{state0}), we can
find  the relation between $|\psi_{n,l} (t)\rangle$ and the stationary
states  $|\psi_{n,l}^s \rangle $. The following relation  holds:
\begin{equation}
|\psi_{n,l}(t) \rangle  \, =\, \hat{R}(t)\,|\psi_{n,l}^s \rangle \, .
\label{psi12}
\end{equation}
\noindent Thus the  vectors $|\psi_{n,l}(t) \rangle$ appearing in the
spectral decomposition of the kernel (\ref{wf1}) have a tight
connection  with $SU(1,1)$ coherent states and, as we shall see soon,
 they describe indeed coherent states which (if expressed
in the $|r,u \rangle$ representation) rotate in their position spread
and/or pulsate in their width (``breathers'').

\vspace{3mm}

\noindent However, before discussing other algebraic properties of the
time--dependent
states $|\psi_{n,l}(t) \rangle$, it is convenient to establish
a connection with the results presented in Refs.\cite{HD1,GV1}.
Let us first denote by $\left\{ |n_A,n_B \rangle \right\}$
the set of eigenstates of $A^{\dagger}A$ and $B^{\dagger}B$. From
(\ref{n121}) follows that $n_A$  and $n_B$ are non--negative
integers. Defining
\begin{equation}
j= \frac{1}{2}(n_A - n_B), \;\;\;\;\; m = \frac{1}{2}(n_A + n_B)\, ,
\end{equation}
\noindent we can label the eigenstates of ${\cal{C}}$ and  $\left(J_3
- \frac{1}{2}\right)$ as $|j,m\rangle $ rather  than $|n_A,n_B
\rangle$. As a result one may write
\begin{equation}
{\cal{C}}|j,m \rangle = j \, |j,m\rangle\, , \;\;\;\; J_3 \,
|j,m\rangle = \left(m + \frac{1}{2} \right)   |j,m\rangle \, ,
\end{equation}
\noindent with two obvious conditions: $|j| = 0, \frac{1}{2}, 1,
\frac{3}{2},  \, \ldots $ and $ m = |j|, |j| +\frac{1}{2}, |j| + 1, \,
\ldots $. The latter is compatible  with (\ref{disc1}).

\vspace{3mm}

\noindent Defining the vacuum state $|0,0\rangle$ as the state
fulfilling   $A\,|0,0\rangle = B\,|0,0\rangle = 0 $, then
the discrete states $|j,m \rangle$  can be  explicitly written as
\begin{eqnarray}
|j,m \rangle = c_{j,m}\, (A^{\dagger})^{ j + m } \, (B^{\dagger})^{ m
- j }\, |0,0\rangle \, ,
\end{eqnarray}
\noindent with
\begin{displaymath}
c_{j,m} = \left[ ( j + m )!\, (m - j)!\right]^{-1/2}\, .
\end{displaymath}
\noindent To relate the eigenstates of $J_2$ with those of $J_3$ above
constructed, we may employ the following relation
\begin{displaymath}
\mbox{exp}(\theta \, J_1)\, J_2 \, \mbox{exp}(-\theta \, J_1) =
-J_2\,\mbox{cos}\theta   - iJ_3\, \mbox{sin}\theta \, ,
\end{displaymath}
\noindent with $\theta \in \mathbb{C}$, to obtain that
\begin{eqnarray}
J_2 = \pm i \, \mbox{exp}\left(\pm \frac{\pi}{2} \, J_1 \right)\, J_3
\;  \mbox{exp}\left(\mp \frac{\pi}{2} \, J_1\right)\, .
\end{eqnarray}
\noindent The former implies that
\begin{eqnarray}
J_2 \, |\Psi^{(\pm)}_{j,m}\rangle &=& \pm i \, \left(m + \frac{1}{2}
\right)  |\Psi^{(\pm)}_{j,m}\rangle\, , \nonumber \\
|\Psi^{(\pm)}_{j,m}\rangle &=& \mbox{exp}\left(\pm \frac{\pi}{2} J_1
\right)|j,m \rangle\, .
\label{state1}
\end{eqnarray}
\noindent Because ${\cal{C}}$ commutes both  with $J_2$ and $J_1$, we
also have
\begin{equation}
{\cal{C}} \, |\Psi^{(\pm)}_{j,m}\rangle = j \,
|\Psi^{(\pm)}_{j,m}\rangle\, .
\label{state2}
\end{equation}
\noindent The relation (\ref{state2}) coincides with the result found
in\cite{HD1}. Similar identity was also realized in
Ref.\cite{GV1}.

\vspace{3mm}

\noindent Comparing (\ref{state1}) and (\ref{state2}) with
(\ref{state0}) we can identify
\begin{equation}
|\psi_{n,l}^s \rangle =\left|\Psi^{(+)}_{n + \frac{l}{2} +
\frac{1}{2},\, \frac{l}{2} - \frac{1}{2}}\right\rangle \, ,
\label{psi1}
\end{equation}
\noindent (note, as $m \geq 0; \; l \geq 1$) or equivalently
\begin{equation}
|\Psi^{(+)}_{j,m} \rangle = \left| \psi_{j-m -1,\, 2m + 1}^s
 \right\rangle \, .
\label{psi2}
\end{equation}
\noindent Similar identification holds for $\left|\Psi^{(-)}_{j,m}
\right\rangle$ ;
\begin{equation}
|\psi_{n,l}^s \rangle = \left|\Psi^{(-)}_{n + \frac{l}{2} +
\frac{1}{2},\, -\frac{l}{2} - \frac{1}{2}} \right\rangle   \, ,
\label{psi3}
\end{equation}
\noindent ($l \leq -1$) or equivalently
\begin{equation}
|\Psi^{(-)}_{j,m} \rangle = \left| \psi_{j+m, \, -2m
-1}^s\right\rangle \, .
\label{psi4}
\end{equation}
\noindent Matching (\ref{TI11}) with (\ref{psi1})--(\ref{psi4}), the
states $|\Psi^{(+)}_{j,m} \rangle$ and $|\Psi^{(-)}_{j,m} \rangle$ can
be related in a simple way, namely
\begin{equation}
{\cal{T}}\,|\Psi^{(+)}_{j,m} \rangle = |\Psi^{(-)}_{j,m} \rangle  =
|\Psi^{(+)}_{j,-(m+1)} \rangle \, ,
\label{TI12}
\end{equation}
\noindent which can be interpreted as a
continuation  of $|\Psi^{(+)}_{j,m} \rangle$
to  negative $m$'s. With (\ref{TI12}) we can check the
consistency of the  inner product defined above. Indeed, using
(\ref{state1}) we have
\begin{eqnarray}
\langle \psi_{n,l}^s| \psi_{n',l'}^s\rangle &=& \left\langle
\Psi^{(-)}_{n+ \frac{l}{2} +\frac{1}{2}, \, \frac{l}{2} -
\frac{1}{2}}\left| \Psi^{(+)}_{n'+\frac{l'}{2}+ \frac{1}{2}, \,
\frac{l'}{2} - \frac{1}{2}} \right. \right\rangle \nonumber \\ &=&
\left\langle n+ \mbox{$\frac{l}{2}$} + \mbox{$\frac{1}{2}$}, \,
\mbox{$\frac{l}{2}$} - \mbox{$\frac{1}{2}$} \left| n'+
\mbox{$\frac{l'}{2}$} +\mbox{$\frac{1}{2}$}, \, \mbox{$\frac{l'}{2}$}
- \mbox{$ \frac{1}{2}$}  \right. \right\rangle \nonumber  \\ &=&
\delta_{nn',\, ll'}\, .
\end{eqnarray}
%
%
%
%
%
%
%
\noindent Now, to understand better the physical nature of
$|\psi_{n,l}(t)\rangle$  let us first note that
$\left(\begin{array} {c} A^{\dagger} \\  B \end{array}\right) $
and $\left(\begin{array} {c} A \\  -B^{\dagger} \end{array}\right)^t
$ are right and left $SU(1,1)$ doublets, respectively. For
instance, under an $SU(1,1)$ rotation
\begin{eqnarray}
e^{i\eta \, J_1}\,  \left(\begin{array}  {c} A^{\dagger} \\ B
\end{array} \right) \, e^{-i\eta \, J_1} &=& \left(\begin{array}{cc}
\mbox{cosh}\frac{\eta}{2} & i\ \mbox{sinh}\frac{\eta}{2} \\ - i\
\mbox{sinh}\frac{\eta}{2}& \mbox{cosh}\frac{\eta}{2}
\end{array} \right)
\,  \left( \begin{array}  {c} A^{\dagger} \\ B \end{array}\right)
\nonumber \\  &=& {\bf{M}} \,\left( \begin{array}  {c} A^{\dagger} \\
B \end{array}\right)  \, .
\label{rot1}
\end{eqnarray}
\noindent The transformation matrix ${\bf{M}}$ is then clearly an
element of the $SU(1,1)$ group as ${\bf{M}}^{\dagger}{\bf{g}} =
{\bf{g}}{\bf{M}}^{-1}\, ;   {\bf{g}} = \mbox{diag} (1,-1)$. Analogous
transformation rules hold also for ``rotations'' with respect to $J_2$
and $J_3$.

\vspace{3mm}

\noindent Thus in terms of the ladder operators $A$ and $B$, the states
$|\Psi^{(\pm)}_{j,m} \rangle $ read
\begin{eqnarray*}
|\Psi^{(\pm)}_{j,m}\rangle  &=&      c_{j,m}\, \left(\frac{A^{\dagger}
\pm  B}{\sqrt{2}}\right)^{ j + m} \, \left(\frac{B^{\dagger} \pm A
}{\sqrt{2}}\right)^{ m -j }\nonumber \\ &\times& \mbox{exp}[\pm\, \pi
/4 \,(A^{\dagger}B^{\dagger} + AB)]\;  |0,0\rangle \, ,
\end{eqnarray*}
\noindent and so
\begin{eqnarray}
|\psi^s_{n,l}\rangle  &=&      \tilde{c}_{n,l}\, (A^{\dagger} + B)^{n}
\, (B^{\dagger} + A )^{|l|-n - 1}\nonumber \\ &\times&
\mbox{exp}[\pi /4 \,(A^{\dagger}B^{\dagger} + AB)]\; |0,0\rangle
\, ,
\label{psis2}
\end{eqnarray}
\noindent with
\begin{displaymath}
\tilde{c}_{n,l} = \left[ n!\, (|l|-n-1)! \, 2^{|l|-1}\right]^{-1/2}\, .
\end{displaymath}
\noindent It follows from (\ref{psi1}) and (\ref{psi3}) that
(\ref{psis2}) is true both for $l\geq 1$ and $l\leq -1$.  Using the
Baker--Campbell--Hausdorff relation one can find that
\begin{eqnarray}
\mbox{exp}[\theta \, J_1]\;  &=& \exp\left[\tan
 \left(\frac{\theta}{2}\right) \, J_+ \right]\, \exp\left[2\lg \left(
 \cos \, \frac{\theta}{2}\right) \, J_3 \right]\nonumber  \\ &\times &
 \exp\left[\tan \left(\frac{\theta}{2}\right) \, J_- \right]\, ,
\label{psis22}
\end{eqnarray}
\noindent (the formula (\ref{psis22}) is an analog of the   Gaussian
decomposition well known from $SO(3)$ group) and so  (\ref{psis2}) can
be recast into a simple form
\begin{eqnarray}
|\psi^s_{n,l}\rangle  &=&      \frac{\tilde{c}_{n,l}}{\sqrt{2}}\,
(A^{\dagger} + B)^{n} \, (B^{\dagger} + A )^{|l|-n -1}\,
\exp(\,J_+)\; |0,0\rangle \nonumber \\ &=& c_{n,l}\,
({\TA^{\dagger}})^{n} \, ({\TB^{\dagger}})^{|l|-n -1} \, | 0
\rangle \rangle \, ,
\label{psis23}
\end{eqnarray}
\noindent with
\begin{eqnarray}
&&{\TA} = \frac{1}{\sqrt{2}}(A - B^{\dagger})\, ,  \;\;\;\;\;
{\TB} =  \frac{1}{\sqrt{2}}(B - A^{\dagger})\, ,\nonumber \\
&&{\TA}^{\dagger} = \frac{1}{\sqrt{2}}(A^{\dagger} + B)\, ,
\;\;\;\;\; {\TB}^{\dagger} =  \frac{1}{\sqrt{2}}(B^{\dagger} +
A)\, .
\end{eqnarray}
\noindent Since the canonical commutation relations are conserved by a
similarity transformation (\ref{rot1}), ${\TA}$ and
${\TA}^{\dagger}$  are new annihilation and creation operators,
respectively (the same holds true for ${\TB}$ and
${\TB}^{\dagger}$), with a new vacuum state $|0\rangle \rangle$
However, because the similarity transformation (\ref{rot1}) for $\eta
=  -i\pi/2$ is not a unitary transformation,  ${\TA}$  and
${\TA}^{\dagger}$ (and ${\TB}$, ${\TB}^{\dagger}$)  are
not  Hermitian conjugates. This should not be surprising: we have already
observed that the state $|\psi^s_{n,l}\rangle$ does transform
under a non--unitary representation of $SU(1,1)$ and  recognized this point
 as the origin of the
 ``anomalous'' behavior of $\psi_{n,l}$ and $\psi^{(*)}_{n,l}$.
However, it should be born in mind that under the inner product above
introduced the ${\TA}$, ${\TA^{\dagger}}$ (and ${\TB}$,
${\TB^{\dagger}}$) are Hermitian conjugates. Indeed, one may
readily  check that,  for instance, $\langle\langle 0 | {\TA} =
\langle\langle {\TA^{\dagger}}\ 0 | $.

\vspace{3mm}

\noindent In connection with Eq.(\ref{psis23}) we should  mention that
$|0\rangle\rangle =1/\sqrt{2} \left[ \exp(\,J_+)\right] |0,0\rangle $
is a two--mode Glauber coherent state\cite{MW1}.  The coherent state
$|0\rangle\rangle$ can be physically  visualized as a boson condensate
or as a new vacuum associated  with ${\TA}$ and ${\TB}$
operators. This suggests to us that  the state $|\psi^s_{n,l}\rangle$
can be interpreted as an excited state of $n$ type--${\TA}$
oscillators, and $(|l|-n-1)$ type--${\TB}$ oscillators with the
coherent state type vacuum  $|0\rangle\rangle =  |\psi^s_{0,-1}
\rangle$.

\vspace{3mm}

\noindent Yet another, interesting interpretation of
$|\psi^s_{n,l}\rangle$   can be obtained if we employ  the $SU(1,1)$
coherent states (i.e., generalized coherent  states\cite{APE1}).
We demonstrate this in Appendix F.

\vspace{3mm}

\noindent Using (\ref{psis23}) it is now simple to interpret
the quantum numbers $n$ and $l$ appearing in the time--dependent states
$|\psi_{n,l}(t)\rangle$. Let us first observe that
\begin{eqnarray}
&&\tilde{\TA}^\dag(t) \equiv  \hat{R}(t)\, {\TA^{\dagger}} \,
\hat{R}^{-1}(t) = a\, {\TA^{\dagger}} - b\, {\TB^{\dagger}}
\nonumber \\
&&\tilde{\TB}^\dag(t) \equiv  \hat{R}(t)\, {\TB^{\dagger}} \,
\hat{R}^{-1}(t) = a^{*} \, {\TB^{\dagger}} - b^{*}\, {\TA^{\dagger}}
\, ,
\label{sp1}
\end{eqnarray}
\noindent (`${*}$' denotes a complex conjugation) where
\begin{eqnarray*}
a &=& \left(1 + \frac{i}{4\Omega}\frac{\dot{\rho}}{\rho} \right)\ \cosh \zeta
- \frac{i}{4\Omega}\frac{\dot{\rho}}{\rho} \ \sinh \zeta \\
b  &=& \left(1 - \frac{i}{4\Omega}\frac{\dot{\rho}}{\rho} \right)\ \sinh \zeta
+ \frac{i}{4\Omega}\frac{\dot{\rho}}{\rho} \ \cosh \zeta \, .
\end{eqnarray*}
and $ |a|^2 - |b|^2 = 1$. In a similar manner, we also find
\begin{eqnarray}
&&\tilde{{\TA}}(t) \equiv  \hat{R}(t)\, {\TA} \,
\hat{R}^{-1}(t) = a^{*}\, {\TA} + b^{*}\, {\TB}
\nonumber \\
&&\tilde{{\TB}}(t) \equiv  \hat{R}(t)\, {\TB} \,
\hat{R}^{-1}(t) = a \, {\TB} + b\, {\TA}
\, .
\label{sp2}
\end{eqnarray}
\noindent Inverting Eqs.(\ref{sp1}) and (\ref{sp2}) we immediately get
the following useful relations:
\begin{eqnarray}
&&\sqrt{2} A = a \left( \tilde{\TB}^\dag +
\tilde{{\TA}} \right) + b^{*} \left( \tilde{\TA}^\dag -
\tilde{{\TB}} \right)\, , \nonumber \\
&&\sqrt{2} A^{\dagger} = a^{*} \left( \tilde{\TA}^\dag -
\tilde{{\TB}} \right) + b \left( \tilde{\TB}^\dag +
\tilde{{\TA}} \right)\, , \nonumber \\
&&\sqrt{2} B = a^{*} \left( \tilde{\TA}^\dag +
\tilde{{\TB}} \right) + b \left( \tilde{\TB}^\dag -
\tilde{{\TA}} \right)\, , \nonumber \\
&&\sqrt{2} B^{\dagger} = a \left( \tilde{\TB}^\dag -
\tilde{{\TA}} \right) + b^{*} \left( \tilde{\TA}^\dag +
\tilde{{\TB}} \right)\, .
\label{pj23}
\end{eqnarray}
\noindent By virtue of unitarity of  $\hat{R}(t)$ the
$\tilde{{\TA}}(t),$
 $\tilde{\TA}^\dag(t)$
(and $\tilde{{\TB}}(t),$ $\tilde{\TB}^\dag(t)$) are
new annihilation and creation operators (albeit
not Hermitian conjugates) with the vacuum state
\begin{displaymath}
\left. \left| [\zeta,\xi ,t] \right\rangle \right\rangle =
\hat{R}(t)\, |0 \rangle \rangle  = \hat{S}(\xi;t) \left. \left|
[\zeta, t] \right\rangle \right\rangle\, .
\end{displaymath}
\noindent Hence, from (\ref{psis23}) we have
\begin{eqnarray}
|\psi_{n,l}(t) \rangle = c_{n,l}\,
(\tilde{\TA}^{\dagger}(t))^{n} \,
(\tilde{\TB}^{\dagger}(t))^{|l|-n -1} \, \left.
\left| [\zeta,\xi,t ]\right\rangle \right\rangle
\, ,
\label{TDCS1}
\end{eqnarray}
\noindent As a matter of fact, the state
\begin{displaymath}
\left. \left| [\zeta, t] \right\rangle \right\rangle = \exp (i\zeta G_A)
\exp (i\zeta G_B)\
|0 \rangle \rangle \, ,
\end{displaymath}
\noindent is also a coherent state (it saturates the uncertainty
relations) called two--mode squeezed  state\cite{MW1} or (from rather
historical reasons)  (two--mode) two--photon coherent state\cite{Y1}.
Term squeeze (or squeezing) coined in\cite{Y1}
reminds that although the dispersions of
canonical variables saturate the uncertainty relations their distribution
over the phase space is distorted (or ``squeezed'') in such a way that the
dispersion of one canonical variable is reduced at the cost of an
increase in the dispersion of the canonically conjugated one.
The concept of squeezing can be extended also to $SU(1,1)$
coherent states\cite{NI1}, however, the actual interpretation is in this
case somehow less clear and states thus obtained do not seem to be
particularly relevant in the present context.

\vspace{3mm}

\noindent The physical meaning of  $\left. \left| [\zeta, \xi, t]
\right\rangle \right\rangle$ can be understood from the
corresponding  dispersions of $\hat{x}_i$ and
$\hat{p}_i$;
\begin{eqnarray}
\left\langle (\Delta \hat{x}_1)^2 \right\rangle &=&
-\frac{\hbar}{2m\Omega} \left[\left\langle \left( A - A^{\dagger}\right)^2
\right\rangle - \left\langle \left( A - A^{\dagger}\right)\right\rangle^2
\right]
 \nonumber \\
&=& \frac{\hbar}{2m\Omega} \ \exp(-2\zeta)\, , \nonumber \\
&& \nonumber \\
\left\langle (\Delta \hat{x}_2)^2 \right\rangle &=&
\left\langle (\Delta \hat{x}_1)^2 \right\rangle \, , \nonumber \\
&& \nonumber \\
\left\langle (\Delta \hat{p}_1)^2 \right\rangle &=&
\frac{\hbar m\Omega }{2} \left[\left\langle \left( A + A^{\dagger}\right)^2
\right\rangle - \left\langle \left( A + A^{\dagger}\right)\right\rangle^2
\right]
 \nonumber \\
&=& \frac{\hbar m \Omega}{2}\left[ \exp (2\zeta) + \left(
\dot{\zeta}/\Omega
\right)^2  \exp (-2\zeta) \right]\, , \nonumber \\
\left\langle (\Delta \hat{p}_2)^2 \right\rangle &=&  \left\langle (\Delta
\hat{p}_1)^2 \right\rangle \, .
\label{cr222}
\end{eqnarray}

\vspace{3mm}

\noindent It is important to realize that (\ref{cr222}) are
obtained using (\ref{pj23}) together with
the modified inner product. So, for example, the relation
\begin{displaymath}
\langle\langle i \tilde{\TA}^{\dagger} [\zeta, \xi, t] | = \{ {\cal{T}}
[i\tilde{\TA}^{\dagger} |[\zeta, \xi, -t] \rangle\rangle ] \}^{\dagger}
= i\langle\langle [\zeta, \xi, t] | \tilde{\TB}^{\dagger}\, ,
\end{displaymath}
\noindent has to be employed. We may easily observe that if  $\rho =
const.$ (i.e.  $\hat{S}(\xi;t) =1$), the dispersions (\ref{cr222})
saturate the Heisenberg uncertainty relations. As a result, the state
$| [\zeta, t] \rangle \rangle$  is indeed the squeezed coherent state
with the {\em squeeze parameter} $\zeta$.  If we evaluate the coherent
state in the position representation we get  a (minimum) wave packet
specified by its half--width (stipulated via the  $\hat{x}_i$
dispersion) and by the mean position (stipulated via mean of
$\hat{x}_i$) \cite{DS1}.  If $\rho \not= const.$ (i.e. $\hat{S}(\xi;t)
\not= 1$) we do not have any more a minimum uncertainty packet -
uncertainty product does not stay $\hbar^2/4$ any more. Apart from the
time dependence gained through $\zeta$ (which would still allow for
the minimum uncertainty wave packet) there is an additional
contribution in the  dispersion of $\hat{p}_i$ as it can be directly
observed from  (\ref{cr222}). So depending on the time behavior of
$\zeta$, the wave packet width now  oscillates and/or spreads and
this, in turn, ``deflects'' the  dispersion of  $\hat{p}_i$ from that
of minimum uncertainty product.

\vspace{3mm}

\noindent It follows that
Eq.(\ref{TDCS1}) allows for a simple physical
explication of $\langle r,u| \psi_{n,l}(t) \rangle$:
the state  $\langle r,u| \psi_{n,l}(t) \rangle$
describes the excited state of $n$ type--$\tilde{{\TA}}$
oscillators, and ($|l|-n-1$) type--$\tilde{{\TB}}$
oscillators with the vacuum state $\langle r,u| [\zeta, \xi, t ]
\rangle\rangle$. The vacuum state itself is represented by the
static wave packet which  pulsates
(and/or spreads) in its width. The excitations then may be understood
as 2D Galilean boosts combined with $SU(1,1)$ rotations.

\section{Geometric phase for Bateman's Dual System}

\subsection{Intermezzo: geometric phase}

\noindent Since in the following an important r{\^ o}le will be played by
geometric phases, we give here a very brief introduction to this subject.

\vspace{3mm}

\noindent One has first to recall that a Hilbert space ${\cal H}$ is a
line bundle over the projective space ${\cal{P}}$, i.e. the equivalence
class of all vectors that differ by a multiplication with a complex
number. We shall denote a generic element of ${\cal{P}}$ as
$|\tilde{\psi}\rangle$.  The inner product on ${\cal H}$ naturally
endows ${\cal{P}}$ with two important geometric structures: a
metric\cite{AA1,JA1}
\begin{equation}
ds^2 = ||d|\tilde{\psi} \rangle ||^{2} - |\langle \tilde{\psi}| d |
\tilde{\psi} \rangle |^{2}\, ,
\label{metr}
\end{equation}
and a $U(1)$ connection (Berry connection\cite{JA1,MN1})
\begin{equation}
{\cal{A}} = i \langle \tilde{\psi} | d |\tilde{\psi} \rangle \, .
\end{equation}
\noindent When a point evolves on ${\cal{P}}$ along a closed loop, say
$\gamma$, the total phase change $\phi_{tot}$ of $|\psi\rangle$ on
${\cal H}$ consist of two contributions: the dynamical part
\begin{equation}
\phi_{dyn} = -\hbar^{-1}\int_{0}^{\tau} dt \,\langle \psi(t)|
{\hat{H}} |\psi(t) \rangle \, ,
\end{equation}
\noindent (with $\tau$ being the time period at which the system
traverses the whole loop $\gamma$), and the geometric part
(Berry--Anandan phase\footnote{It would be perhaps more correct to
call the geometric phase presented in this paper as Anandan's
phase\cite{JA1} or Aharonov--Anandan's phase\cite{AA1}. However, due
to a historical reasons, the Berry phase\cite{MVB11} is usually taken as
a synonym for any cyclic geometric phase.
We feel that the name Berry--Anandan phase is a suitable compromise between
rigor and tradition.})
\begin{eqnarray}
e^{i\phi_{BA}} &=& e^{i\phi_{tot} - i \phi_{dyn}}\nonumber \\   &=&
\langle \psi(0)|\psi(\tau)\rangle \,
\mbox{exp}\left(i\int_{0}^{\tau}dt \, \langle
\psi(t)|i\frac{d}{dt}|\psi(t) \rangle \right) \nonumber \\
&=&\mbox{exp}\left(i\int_{0}^{\tau}dt \, \langle \tilde{\psi}(t)|i
\frac{d}{dt}  | \tilde{\psi}(t)\rangle \right)\nonumber \\ &=&
\mbox{exp}\left(i\int_{\gamma} {\cal{A}}\right)\, .
\label{berryp1}
\end{eqnarray}
\noindent So the Berry--Anandan phase $\phi_{BA}$ may be geometrically
understood as  (an)holonomy with respect to the natural (Berry's)
connection on the projective space ${\cal{P}}$ \cite{AA1}. It is needless
to say that all above relations are meant to be valid only for states
$|\psi(t) \rangle$ being normalized to unity. In general, corresponding
division of the state norm must be invoked. As our states
$\psi_{n,l}(r,u,t)$ are not normalized, this should be kept
in mind in the following where, for clarity of notation,
we shall omit  this normalization factor.

\vspace{3mm}

\noindent Actually, the geometric phase can also be defined for open
paths.  In this case the phase is usually referred to as
Pancharatnam's phase\cite{JS1}.  The trick is  that any path on
${\cal{P}}$ can be closed by joining endpoints with a geodesic
constructed with respect to the metric (\ref{metr}). The geometric
phase of the loop thus constructed is then defined to be equal to the
geometric phase associated to the open path. So, if $\gamma_o$  is an
open path on ${\cal{P}}$  and $\gamma_{g}$ is the corresponding
geodesic on ${\cal{P}}$, then the associated Pancharatnam phase
$\phi_P$ reads
\begin{eqnarray}
e^{i\phi_P} &=& \mbox{exp}\left(i\int_{\gamma_o + \gamma_g}\,
{\cal{A}} \right) = \mbox{exp}\left(i\int_{\gamma_o}\, {\cal{A}}
\right) \nonumber \\ &=& \mbox{exp}\left( i \int_{t_i}^{t_f}dt\,
\langle \tilde{\psi}(t)|  i\frac{d}{dt} | \tilde{\psi}(t) \rangle
\right) \nonumber \\ &=&  \langle \psi(t_i)|\psi(t_f)\rangle
\,\mbox{exp}\left( i  \int_{t_i}^{t_f}dt\, \langle \psi(t)|
i\frac{d}{dt} | \psi(t) \rangle \right)\, .  \lab{pan1}
\label{berryp2}
\end{eqnarray}
\noindent Here we have used the fact that parallel transport along a
geodesic does not bring any anholonomy. Note that $\phi_P$ is well
defined only if the endpoints are not orthogonal. It should be also
clear that  $\phi_P$ is defined only modulo $2\pi$.

\subsection{Exact geometric phase in Bateman's system}

\noindent To find the geometric phase for Bateman's system we firstly
compute the dynamical phase
\begin{eqnarray*}
\phi_{dyn} &=& -\int_{t_i}^{t_f} dt \, \langle
\psi_{n,l}(t)|i\frac{d}{dt}  | \psi_{n,l}(t) \rangle \nonumber \\ &=&
-\hbar^{-1}\int_{t_i}^{t_f} dt \, \langle \psi_{n,l}(t)|{\hat{H}}  |
\psi_{n,l}(t) \rangle \, .
\end{eqnarray*}
\noindent Using (\ref{wave}), (\ref{qn1}) and (\ref{C1}) we find that
\begin{eqnarray}
\phi_{dyn} &=&    -(2n + l + 1)\int_{t_i}^{t_f}dt\,
\frac{\sqrt{W}}{2\rho}\, \left( \frac{{\dot{\rho}}^2}{4W} +
\frac{\Omega^2 \rho^2}{W} + 1 \right) \nonumber \\ &&+\,  i\,\Gamma
l\, (t_f -t_i) \, .
\label{qn2}
\end{eqnarray}
\noindent In order to proceed further it is convenient to define the
complex  number %
\begin{displaymath}
z(t)  = i\,\sqrt{\frac{V(t)}{\rho(t)}} + \sqrt{1 -\frac{V(t)}{\rho(t)}
}\, .
\end{displaymath}
\noindent Notice that $z{\bar{z}} =1$. When evolving in the time
interval $(t_i,t_f) $, $z(t)$  traverses a curve $\gamma$ in the
Gaussian plane. The index of the curve $\gamma$ (i.e. the number of
revolutions around the origin)  is then defined as\cite{VIA1}:
\begin{equation}
\mbox{ind}\, \gamma = \frac{1}{2\pi i} \oint_{\gamma} \frac{dz}{z}\, .
\label{ind1}
\end{equation}
\noindent Consequently the dynamic phase can be rewritten in the form
\begin{eqnarray}
\phi_{dyn} &=& - (2n + l +1)\int_{t_i}^{t_f}dt \left(
\frac{{\dot{\rho}}^2}{8\rho \sqrt{W}} + \frac{\Omega^2 \rho}{2
\sqrt{W}}\right) \nonumber \\  && \nonumber \\ &&-\, (2n + l +
1)(\pi\, \mbox{ind}\, \gamma) + i\, \Gamma l\,(t_f -  t_i)\, .
\label{qn12}
\end{eqnarray}
\noindent  Using (\ref{ind1}) we may write also  the total phase in a
fairly compact form, indeed
\begin{eqnarray}
\phi_{tot} &=& arg\left\{\langle \psi_{n,l}(t_i)|\psi_{n,l}(t_f)
\rangle \right\}\nonumber \\ &=& -(2n + l+1)\,\left(
\mbox{arcsin}\sqrt{\frac{V(t_f)}{\rho(t_f)}} -
\mbox{arcsin}\sqrt{\frac{V(t_i)}{\rho(t_i)}}\right)\nonumber \\ &&  -
\;\frac{\pi}{2}\, n_{i,f} + \; i \Gamma l (t_f -t_i) \nonumber \\ &=&
-(2n + l+1)(2\pi\, \mbox{ind}\, \gamma\,) -\frac{\pi}{2}\, n_{i,f} + i
\Gamma l (t_f -t_i) \, , \nonumber \\
\label{tot1}
\end{eqnarray}
\noindent With $n_{i,f}$ being the Morse index of the classical
trajectory running between ${\bf{x}}_i$ and ${\bf{x}}_f$.

\vspace{3mm}

\noindent At this stage a remark should be added.  The fact that we
get an imaginary piece both in $\phi_{tot}$ and $\phi_{dyn}$ should
not be surprising because we work with the modified inner product.
Notice that the ``troublesome"
contribution in $\phi_{tot}$ and $\phi_{dyn}$ correctly flips sign if
one passes from $\psi(\ldots)$ to $\psi^{(*)}(\ldots)$.

\vspace{3mm}

\noindent Substituting (\ref{qn12}) and (\ref{tot1}) into the equation
(\ref{pan1}),   we  get  Pancharatnam's   phase
\begin{eqnarray}
\phi_P &=& (2n + l + 1)  \int_{t_i}^{t_f}dt  \left(
\frac{{\dot{\rho}}^2}{8\rho       \sqrt{W}}      +      \frac{\Omega^2
\rho}{2\sqrt{W}}\right)  \nonumber \\    &-& (2n + l + 1) \left( \pi\,
{\mbox{ind}}\,  \gamma  \right) \nonumber \\ &-&  \frac{\pi}{2}\,
n_{i,f} \, .
\label{panch}
\end{eqnarray}
\noindent Note that because $\rho, W$ and $V$ are solely constructed
out of   solutions of the classical equations of motion, $\phi_P$ is
manifestly $\hbar$ independent.

\vspace{3mm}

\noindent The meaning of the terms on the RHS of
(\ref{panch})  can be easily understood.  Let us assemble the first two
pieces in Eq.(\ref{panch}) together. Using the fact that
\begin{displaymath}
\rho = \frac{\sqrt{W}}{\Omega}\, \exp (-2 \zeta)\, ,
\end{displaymath}
\noindent (see explanation below Eq.(\ref{C1})) together with
(\ref{rho1}) we obtain
\begin{eqnarray}
\phi_P  &=& (2n + l + 1) \,
\frac{\Omega}{2}\, \int_{t_i}^{t_f} dt \left(  \exp (-2\zeta)
+  \exp
(2 \zeta) \right.
\nonumber \\
&+&  \left. \exp (-2\zeta) \, (\dot{\zeta}/\Omega) \right) + \frac{\pi}{2}\,
n_{i,f}\nonumber \\
&& \nonumber \\
&=& (2n + l + 1) \,\int_{t_i}^{t_f} dt \left( \frac{\left\langle
(\Delta \hat{p}_i )^2 \right\rangle}{\hbar m}
+ \frac{m\Omega \left\langle (\Delta \hat{x}_i )^2 \right\rangle }{\hbar}\,
\right)
\nonumber \\
&+& \frac{\pi}{2}\, n_{i,f}\, .
\label{PP11}
\end{eqnarray}
\noindent The index ``$i$'' is either $1$ or $2$ (it really does not matter
as dispersions are symmetric, one may write also the symmetrized version with
the prefactor $\frac{1}{2}$, however, some
care should be taken as we do not have
Euclidean scalar product and so, for instance, $(\hat{p}_2)^2 = - \hat{p}_2
\hat{p}^2 $). It is important to recognize that $\langle \ldots \rangle$
in (\ref{PP11}) represents the mean value with respect to the {\em ground}
state.

\vspace{3mm}

\noindent Thus $\phi_P$ is a collection of three contributions;
overall ground--state fluctuations of $\hat{p}$ and $\hat{x}$ gathered
during the time period $t_f -t_i$ and the Morse index. While the first two
are basic characteristics of the ground--state wave packet,
the Morse index contribution, on the other hand, reflects the
geometrical features  of the path traversed by the ground--state wave
packet in the configuration--space.  As explained before, its presence is
inevitable for providing a correct analytical continuation of the
Feynman--Hibbs kernel prescription around the focal points.

\vspace{3mm}

\noindent The above considerations show that the properties of $\phi_P$ are
basically  encoded in the structure and in the time dependence of the
ground state.  This intertwining of the ground  state with geometric
phase will be of a crucial importance in the following.

\vspace{3mm}

\noindent An interesting question which one can raise in the present
context is how the non--abelian (Wilczek--Zee) geometric
phase\cite{WZ1}  looks in Bateman's system and to what extent it
influences the presented  results. This is definitely a challenging
task as there does not exist at present  any formulation of
non--abelian geometric phases outside of the scope of adiabatic
approximation (i.e. geometric phases pioneered by Berry). As the
geometric phases  (\ref{berryp1}), (\ref{berryp2}) and (\ref{panch})
are rather  Aharonov--Anandan type, such an extension would be of a
particular interest. We intend to investigate this question in the
future work.

%

\section{The ground state of the 1D linear harmonic oscillator}
\subsection{Berry--Anandan phase}

\noindent An interesting implication of (\ref{panch}) arises
when we turn our attention to the special case of  $l=
-\frac{1}{2}$. From Section III we know that such a choice corresponds
to the 1D l.h.o.. In addition, from  (\ref{wave})
and (\ref{wave4}) we see that the following relation between the  1D
l.h.o. wave function  $\psi_n^{lho} (r,t)  \equiv
\psi_{n, - \frac{1}{2}}(r,t)$ and Bateman's system wave  function
$\psi_{n, l}(r,u,t)$ holds:
\begin{equation}
\psi_n^{lho} (r,t) = \sqrt{\pi \,r}\,\psi_{n, - \frac{1}{2}} (r,
-\Gamma t + \beta/2, t)\, .
\label{rel3}
\end{equation}
\noindent The important point about $\psi_n^{lho} (r,t)$ is that it
is  constructed exclusively from the fundamental system of solutions
corresponding to Bateman's dual system, and not, as one could expect,
from fundamental system of solutions of the 1D l.h.o..
In addition, it should be noted that the value  $u =
-\Gamma t + \beta/2$ entering (\ref{rel3}) is nothing but the solution
of the {\em classical} equations  of the motion: ${\dot{u}}= \partial
H/\partial p_u$, ${\dot{p}}_u = - \partial H/\partial u$  with the
classical $p_u = J_2 = 0$ and Bateman's  dual Hamiltonian $H$. One may
naturally wonder then whether some measurable information about the
original system could be tracked down in $\psi_n^{lho} (r,t)$.
We shall see that this is indeed the case:
a memory of the underlying Bateman's system is
imprinted in  the ground  state energy of the reduced system.

\vspace{3mm}

\noindent Let us look at the
geometric phase of the 1D l.h.o.  ``inherited''
through the reduction  (\ref{rel3}). Being  $\tau$
the mutual period of both $\rho$ and $V$, then the
following chain of reasonings holds:
\begin{eqnarray}
&&\psi_{n}^{lho}(r,\tau ) = \sqrt{\pi \, r} \; \psi_{n,
-\frac{1}{2}}(r,  u, \tau )|_{u= \beta/2-\tau \Gamma }\nonumber \\
&&\mbox{\hspace{2mm}} = \sqrt{\pi \, r} \, \left( e^{i\phi_{tot}}\,
\psi_{n, -\frac{1}{2}}(r,u,  0)\right)|_{u= \beta/2-\tau \Gamma
}\nonumber \\ &&\mbox{\hspace{2mm}}= \sqrt{\pi \, r} \, \left(
e^{i\left[\phi_{tot} -  \tau\Gamma  p_{u}/\hbar\right]}\, \psi_{n,
-\frac{1}{2}}(r,u + \tau \Gamma,  0)\right)|_{u= \beta/2-\tau \Gamma
}\nonumber \\
&&\mbox{\hspace{2mm}}=  e^{i\left[\phi_{BA} +  \frac{1}{\hbar}\,
\int_{0}^{\tau}\langle  \psi_n^{lho}(t)| {\hat{H}}_{-\frac{1}{2}} |
\psi^{lho}_n (t) \rangle\, dt  \right]}\; \psi_n^{lho}(r,0)\, .
\label{BA3}
\end{eqnarray}
\noindent On the other hand, because of the periodicity of
 $\rho$ and $V$, the wave function $\psi_{n}^{lho}(r,t)$ must be
$\tau$--periodic  as well. This implies that
$\psi_{n}^{lho}(r,0) =  e^{i2\pi m } \,\psi_{n}^{lho}(r,\tau)$, where
$m$ is an arbitrary integer.   Using the fact that $\phi_{BA}$ is
defined modulo $2\pi$  we can write
\begin{equation}
\int_{0}^{\tau} \langle \psi_n^{lho}(t)| {\hat{H}}_{-\frac{1}{2}} |
\psi^{lho}_n  (t)  \rangle \, dt =  \hbar\, \left(2\pi n -
\phi_{BA}\right)\, .
\label{keye}
\end{equation}
\noindent In particular, if $\psi_n^{lho}(r,t)$ are eigenstates of
${\hat{H}}_{-\frac{1}{2}}$, then we get the
quantized  energy spectrum:
\begin{equation}
\mbox{E}^{lho}_n = \frac{\hbar}{\tau}  \left(2\pi n
-\phi_{AB}\right)\, .
\label{keye1}
\end{equation}
\noindent In the usual semiclassical treatment the presence of the
Berry--Anandan phase modifies the energy spectrum via  the
Bohr--Sommerfeld quantization condition\cite{RGL3,MV1}. In the case of
the simple 1D l.h.o.
(not the one obtained from the Bateman's system after reduction),
 the Berry--Anandan phase
materializes only due to the Morse index contribution,
i.e. $\phi_{BA}= -\pi\,n_{a,b}/2$.  When $(t_{b}-t_a) =
\tau = 2\pi/\Omega$, the Morse index  is simple\cite{HK1}: $n_{a,b}=2$
and one recovers the standard relation:
\begin{equation}
{\tilde{\mbox{E}}}^{lho}_n = \hbar \Omega \left(n +
\frac{1}{2}\right)\, .
\label{keye2}
\end{equation}
\noindent In this respect the result (\ref{keye1}) might seem rather
peculiar,  especially in the case when $\phi_{BA} \not= -\pi$.  The
fact that $\phi_{BA}$ can indeed be different from $-\pi$ will be
explicitly illustrated in the following subsection. However,  the
basic reason for this to happen is not difficult to understand:
the Morse index of the underlying Bateman's system (at $u =
-\Gamma \tau $ ) is not necessarily equal to the Morse index of the
corresponding 1D l.h.o. at the same elapsed time
$\tau$.

\vspace{3mm}

\noindent Let us add two more comments at the end. Firstly,
the foregoing analysis can be naturally extended on the case
$l = + \frac{1}{2}$, but instead of doing that we may directly
observe from
(\ref{wave4}) that
\begin{equation}
\tilde{\psi}_{n}^{lho}(r,t) \equiv \psi_{n,\frac{1}{2}}(r,t)
= \psi_{n+\frac{1}{2}, -\frac{1}{2}}(r,t)\, ,
\end{equation}
\noindent so the 1D l.h.o. states obtained from
$\psi_{n,\frac{1}{2}}(r,t)$ describe the higher energy states than
$\psi_{n}^{lho}(r,t)$, and thus, for instance, the ground state comes
entirely from the  $l= - \frac{1}{2}$ case. Secondly, it may happen
that $\rho$ and $V$ can have more than one common  period.  Such
periods can give rise to (generally)  different (not mod$(2\pi)$)
Berry--Anandan phases. This situation is  fairly standard in many
systems (see e.g., \cite{DYS1}) and the corresponding phases have as a
rule different physical consequences.

\subsection{Practical example - stationary states }

\noindent To elucidate the previous analysis, we consider here an
explicit  example in which the following fundamental system of
solutions is chosen:
\begin{eqnarray}
u_{1}^{1}(t)&=& \sqrt{2} \cos{(\Om t)} \;\cosh{(\Ga t)}\, ,\nonumber  \\
u^{1}_{2}(t)&=& -\sqrt{2} \cos{(\Om t)}\; \sinh{(\Ga t)}\, ,\nonumber  \\
u^{2}_{1}(t)&=& \sqrt{2} \cos{(\Om t)}\; \sinh{(\Ga t)}\, ,\nonumber \\
u_{2}^{2}(t)&=& -\sqrt{2}  \cos{(\Om t)}\; \cosh{(\Ga t)}\, ,\nonumber \\
v_{1}^{1}(t)&=& \sqrt{2} \sin{(\Om t)}\; \cosh{(\Ga t)}\, ,\nonumber \\
v^{1}_{2}(t)&=& -\sqrt{2} \sin{(\Om t)} \;\sinh{(\Ga t)}\, ,\nonumber \\
v^{2}_{1}(t)&=& \sqrt{2} \sin{(\Om t)}\; \sinh{(\Ga t)}\, ,\nonumber \\
v_{2}^{2}(t)&=& -\sqrt{2} \sin{(\Om t)} \;\cosh{(\Ga t)}\, .
\label{FS1}
\end{eqnarray}

\noindent The Wronskian is
\begin{equation}
W = \lf| \ba{cccc} \sqrt{2} &0&0&0 \\ 0&-\sqrt{2} &0&0 \\
0&\sqrt{2}\Ga &\sqrt{2}\Om&0 \\ -\sqrt{2}\Ga &0&0&-\sqrt{2}\Om \ea\ri|
= 4 \Om^2 \, ,
\label{FS2}
\end{equation}
\noindent and the determinant $D$
\begin{displaymath}
D= 4 \sin^2{\Om(t_b -t_a)}\, .
\end{displaymath}
\noindent As a result we have
\bea\non B_1^1(t)&=&4 \,\cosh[\Ga (t-t_a)]\, \sin[\Om
(t_b-t)]\,\sin[\Om (t_b-t_a)]\, , \\ \non
B_3^1(t)&=&4 \,\cosh[\Ga
(t_b-t)]\, \sin[\Om (t-t_a)]\,\sin[\Om (t_b-t_a)]\, , \\ \non
B_1^2(t)&=&4
\,\sinh[\Ga (t_a-t)]\, \sin[\Om (t_b-t)]\,\sin[\Om (t_b-t_a)]\, .  \eea

\noindent The classical action has the form
\begin{eqnarray}
S_{cl}&=& \frac{m \Om}{2 \sin{[\Om (t_b-t_a)]}} \lf\{  (r_a^2 + r_b^2)
\cos{[\Om (t_b-t_a)]} \ri.  \nonumber \\ &-&\lf.  2 r_a r_b
\cosh{[u_b-u_a- \Ga (t_b -t_a)]} \ri\}\, ,
\label{FS6}
\end{eqnarray}
\noindent and we find that the fluctuation factor reads
\begin{equation}
F[t_a, t_b]= \frac{m}{2 \pi \hbar} \,\frac{\Om}{|\sin{[\Om
(t_b-t_a)]}|}\, .
\label{FS5}
\end{equation}
\noindent Eqs.(\ref{FS6}) and (\ref{FS5}) lead to the kernel:
\begin{eqnarray}
&&\langle r_{b}, u_{b}; t_{b}| r_{a}, u_{a}; t_{a} \rangle =
\frac{m}{2 \pi \hbar} \,\frac{\Om}{|\sin{[\Om (t_b-t_a)]}|}
\nonumber \\
&&\, \times \, \exp\left[\frac{i m \Om}{2 \hbar \sin{(\Om (t_b-t_a))}}
\lf\{  (r_a^2
+ r_b^2) \cos{(\Om (t_b-t_a))} \ri. \ri.\nonumber \\
&& \nonumber \\
&& \, - \lf.  \, 2 r_a r_b \cosh{(\Delta u- \Ga (t_b
-t_a))} \ri\}\mbox{\huge{]}}\, .
\label{FS7}
\end{eqnarray}
\noindent Note that the kernel is indeed independent of the
fundamental  system of solutions. One may check that the kernel
(\ref{FS7}) satisfies the  time--dependent Schr{\"o}dinger equation
(\ref{Ham1}).

\vspace{3mm}

\noindent We now rewrite the kernel applying the expansion
(\ref{ker11}). Remembering that there is an absolute value in
(\ref{FS7}) and employing the fact that
\bea\non V(t) & =& 2
\sin^2(\Om t)\, , \\ \non
\rho(t)&=& 2\, , \\ \non
b(t) &=& e^{-2 i \Om t}\,
, \eea
\noindent we get
\bea\non &&\langle r_{b}, u_{b}; t_{b}| r_{a}, u_{a}; t_a\rangle =
\frac{i}{\pi} \,\sum_{n,l}
\frac{n!}{\Ga(n+l+1)}\,\lf(\frac{m\Om}{\hbar}\ri)^{l +1}     \\  \non
&& \times  \,L_n^l\lf(\frac{m \Om}{\hbar} r_a^2\ri)\,
L_n^l\left(\frac{m \Om}{\hbar} r_b^2\right)\, (r_a r_b)^l \,
e^{-\frac{m\Om}{2 \hbar}(r_a^2 +r_b^2) }  \nonumber \\  && \times
\,e^{-i \Om (2n + l + 1)(t_b-t_a)}   \, e^{-l \Delta u - l \Ga (t_b
-t_a) }\, .  \eea
\noindent The explicit form of the wave function is then
\begin{eqnarray}
\psi_{n,l}(r,u,t) &=& \sqrt{\frac{n!}{ \pi\Ga(n+l+1)}} \,
\lf(\frac{m\sqrt{\Om}}{ \hbar}\ri)^{l+\frac{1}{2}} \, r^l\,
e^{-\frac{m\Om}{2 \hbar}r^2} \nonumber \\ &\times& L_n^l\left(\frac{m
\Om}{\hbar}r^2\right) \, e^{-l(u +\Ga t)}\, e^{-i\Om (2n +l +1)t}\, .
\label{wf4}
\end{eqnarray}
\noindent The radial wave function is on the other hand:
\begin{eqnarray}
\psi_{n,l}(r,t) &=&  \sqrt{\frac{n!}{\pi \Ga(n+l+1)}} \,
\lf(\frac{m\sqrt{\Om}}{\hbar}\ri)^{l+\frac{1}{2}} \,
r^{l+\frac{1}{2}}\, e^{-\frac{m\Om}{2 \hbar}r^2} \, \nonumber \\
&\times &\, L_n^l\left(\frac{m \Om}{\hbar}r^2\right) \, e^{-i\Om (2n
+l +1)t} \, .
\end{eqnarray}
\noindent Note that $\psi_{n,l}(r,t)$ is an eigenstate of
${\hat{H}}_{l}$. In passing it is also interesting to consider the reduced
wave  function $|\psi\ran_{H_\I}$ which is obtained from the full one
through the formula $|\psi\ran_{H_\I}=\exp[-  \Ga t \pa_u]
|\psi\ran_{H} $. This amounts to substitute $u$ with $(u-\Ga t)$ into
the total wave function (\ref{wf4}), which then becomes
\begin{eqnarray}
&&\psi_{n,l}(r,u-\Ga t ,t) =   \sqrt{\frac{n!}{\pi \Ga(n+l+1)}} \,
\lf(\frac{m\sqrt{\Om}}{ \hbar}\ri)^{l+\frac{1}{2}}\nonumber \\
&&\mbox{\hspace{5mm}}\times \, r^l\, e^{-\frac{m\Om}{2 \hbar}r^2}
\ L_n^l\left(\frac{m
\Om}{\hbar}r^2\right)\, e^{-l u }\, e^{-i\Om (2n
+l +1)t}\, .
\end{eqnarray}
It is easy to verify that it satisfies the reduced Schr\"odinger
equation:
\bea i \pa_t \, \psi_{n,l}(r,u-\Ga t ,t) \, =\, 2 \Om {\cal C}\,
\psi_{n,l}(r,u-\Ga t ,t)\, ,  \eea
\noindent This result will be particularly important in
the following paper of this series.
%
%
%

\vspace{3mm}

\begin{figure}[h]
\epsfxsize=6.5cm
\centerline{\epsffile{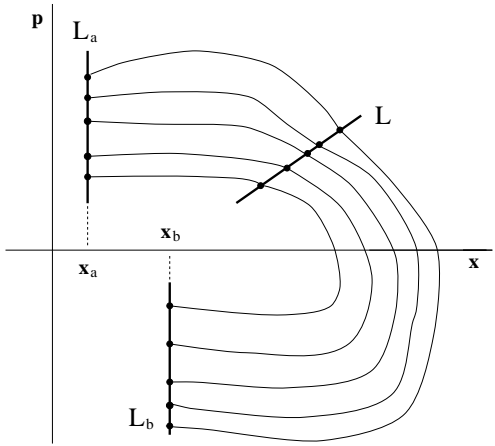}}
\vspace{4mm}
\caption{{\em
Evolution of a Lagrangian manifold schematically depicted at
three distinct times. Note that due to the initial condition} (4)
{\em we have} $\mbox{L}_a = \{{\bf{x}}_a, {\bf{p}}_a = anything\}$,
{\em note also that
the Lagrangian manifold} L$_b$ {\em is
responsible for a caustic at ${\bf{x}}_b$.
Five possible orbits are shown.}}
\label{fig1}
\end{figure}

\noindent Because $V$ is periodic with {\em fundamental period}  $\tau =
\pi/\Omega$ and because $\psi_{n,-\frac{1}{2}}$ is an eigenstate of
${\hat{H}}_{-\frac{1}{2}}$, the energy spectrum of the related 1D
l.h.o. is done by the prescription
(\ref{keye1}). The corresponding ground--state energy can be
calculated from (\ref{panch}) and (\ref{keye1}). We obtain
\begin{equation}
E_{0}^{lho} = - \hbar \, \frac{\phi_{BA}}{\tau} = \hbar \pi
\, \frac{n_{a,b}}{2\tau} \, .
\label{gs12}
\end{equation}
\noindent We see therefore that the fundamental system (\ref{FS1})
reflects the dynamics of the underlying Bateman's dual system in
$E_{0}^{lho}$ only via the Morse index $n_{a,b}$. To find $n_{a,b}$,
we first define a Lagrangian
manifold\cite{VPM1,RGL1}: $\mbox{L} = \{ {\bf x}, \,{\bf p} = \partial
S_{cl}/ \partial{\bf x}\} $, where $S_{cl}$ is the action taken as a
function of the end point ${\bf x}$ ($= {\bf x}_b$). For quadratic
actions the Lagrangian manifold is clearly $n$--dimensional plane in
the $2n$--dimensional phase space.   Because the starting point of our
analysis was the (configuration--space) kernel    we are primarily
interested in orbits for which  the initial and final positions  are
given. Of course, if the initial position and momenta were given, then
we would have a unique orbit, but since instead we have initial and
final positions, it is not clear that any such orbit exists,  and if
does, whether it is unique. Let us therefore consider   the time
evolution of L. As time progresses,  the Lagrangian manifold evolves
(foliates) the phase space\footnote{Due to its very definition,
Lagrangian manifolds  transform under canonical transformations into
other Lagrangian  manifolds\cite{VPM1}.  So namely Lagrangian manifolds
evolve into other Lagrangian manifolds  under (Hamiltonian) time
evolution.}. If we denote a point of the   initial--time Lagrangian
manifold as ${\bf{z}}_a = ({\bf{p}}_a,{\bf{x}}_a )$  (note that set of
all such point forms the Cauchy data for the  Hamiltonian dynamics)
then due to quadratic nature of $H$  the final--time point
${\bf{z}}_b$ is related with ${\bf{z}}_a$ via linear canonical
transformation (symplectic matrix): ${\bf{z}}_b  =  {\mathbb{S}}
{\bf{z}}_a\,$, where ${\mathbb{S}}= {\mathbb{S}}(t_a,t_b)$.  Taking
only the ${\bf{x}}$ part of ${\bf{z}}_b$ we may write
\begin{equation}
{\bf{x}}_b = {\mathbb{S}}_1 {\bf{p}}_a + {\mathbb{S}}_2 {\bf{x}}_a
\, \; \; \Leftrightarrow  \;\; {\mathbb{S}}_1 {\bf{p}}_a = -
{\mathbb{S}}_2 {\bf{x}}_a + {\bf{x}}_b \, .
\label{sympl1}
\end{equation}
\noindent Clearly, if ${\mathbb{S}}_1(t_a,t_b)$ were invertible
(i.e., if $\det({\mathbb{S}}_1)\not= 0 $)  then for given points
${\bf{x}}_b$ and ${\bf{x}}_a$ would exist only one ${\bf{p}}_a$ and
consequently only one classical orbit  would run between ${\bf{x}}_b$
and ${\bf{x}}_b$.  If however $\det({\mathbb{S}}_1) = 0 $,  then
 either none or infinitely many
solutions may be obtained,  depending on the ranks of ${\mathbb{S}}_1$
and the corresponding augment matrix.   So if  $\det({\mathbb{S}}_1) =
0$ and ${\bf{x}}_a$ and ${\bf{x}}_b$ are such that Eq.(\ref{sympl1})
is not satisfied, then there are no orbits arriving at
${\bf{x}}_b$. If, however,  ${\bf{x}}_a$ and  ${\bf{x}}_b$ satisfy
Eq.(\ref{sympl1}), then there is an $n$-dimensional  infinity of initial
momenta which maps onto ${\bf{x}}_b$ and thus  density of particles is
infinite at ${\bf{x}}_b$ - orbits are focused  in the configuration
space (the situation is schematically depicted in Fig.[\ref{fig1}]).

\noindent It is not difficult to see
that the precarious points ${\bf{x}}_b$ in which $\det({\mathbb{S}}_1)
= 0$ are precisely the focal (conjugate) points. Indeed,
\begin{displaymath}
0 = \det({\mathbb{S}}_1) = {\det}_2 \left(\frac{\partial
{\bf{x}}_b^{\alpha}}{\partial {\bf{p}}_a^{\beta}} \right) = {\det}_2
\left[ \left( \frac{\partial^2 S_{cl}}{\partial  {\bf{x}}_a^{\alpha}
{\bf{x}}_b^{\beta} }   \right)^{-1}\right]\, .
\end{displaymath}

\noindent When passing through a caustic, $\det({\mathbb{S}}_1)$  may
change the sign depending on the rank of
$\det({\mathbb{S}}_1(t_a,t_b))$ at the caustic.  The caustic is said to
have multiplicity  $k$ if the rank of $\det({\mathbb{S}}_1(t_a,t_b))$
is $k$. The change of sign will directly influence the
form of the  fluctuation factor $F[t_a,t_b]$  (which is basically
the square root of  $1/\det[{\mathbb{S}}_1(t_a,t_b)]$) as the correct
branch cut must be chosen. The phase of  $F[t_a,t_b]$ can be
consistently prescribed demanding continuity in the kernel\cite{VPM1}.
It turns out that a phase factor $\exp(-ik\pi/2)$ must appear when
passing a caustic of multiplicity $k$.  As more caustics are passed
along an orbit, the phase factor will accumulate.  The Morse index
$n_{a,b}$ appearing in (\ref{ker112}) then simply counts  caustics
(including their multiplicity) encountered by an orbit passing from
the initial to the final Lagrangian  manifold.

\vspace{3mm}

\begin{figure}[h]
 \epsfxsize=8.5cm \centerline{\epsffile{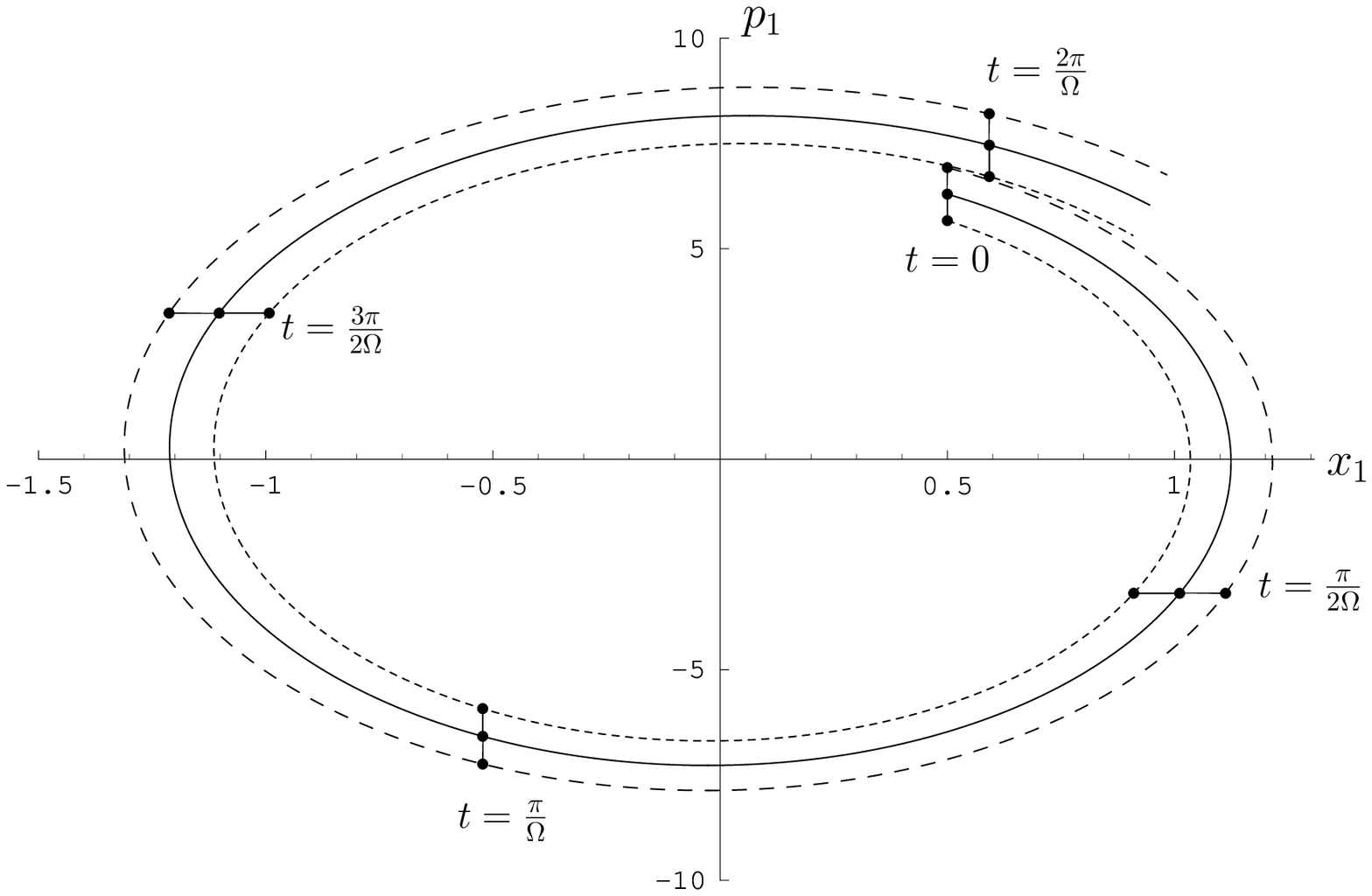}}
\vspace{5mm}
\caption{\em
Evolution of a Lagrangian manifold in $(x_1,p_1)$ phase
space. Three orbits with different values of ${\bf{p}}_a$ are plotted.
The Lagrangian
manifold is shown at five different times.
Note that times $t = \pi/\Omega$ and $t = 2\pi/\Omega$
correspond to the conjugate times. This
and the following plots are done for $\ka=40$, $\ga=1.2$, $m=1$.}
\label{fig2}
\end{figure}

\begin{figure}[h]
 \epsfxsize=8.5cm
\centerline{\epsffile{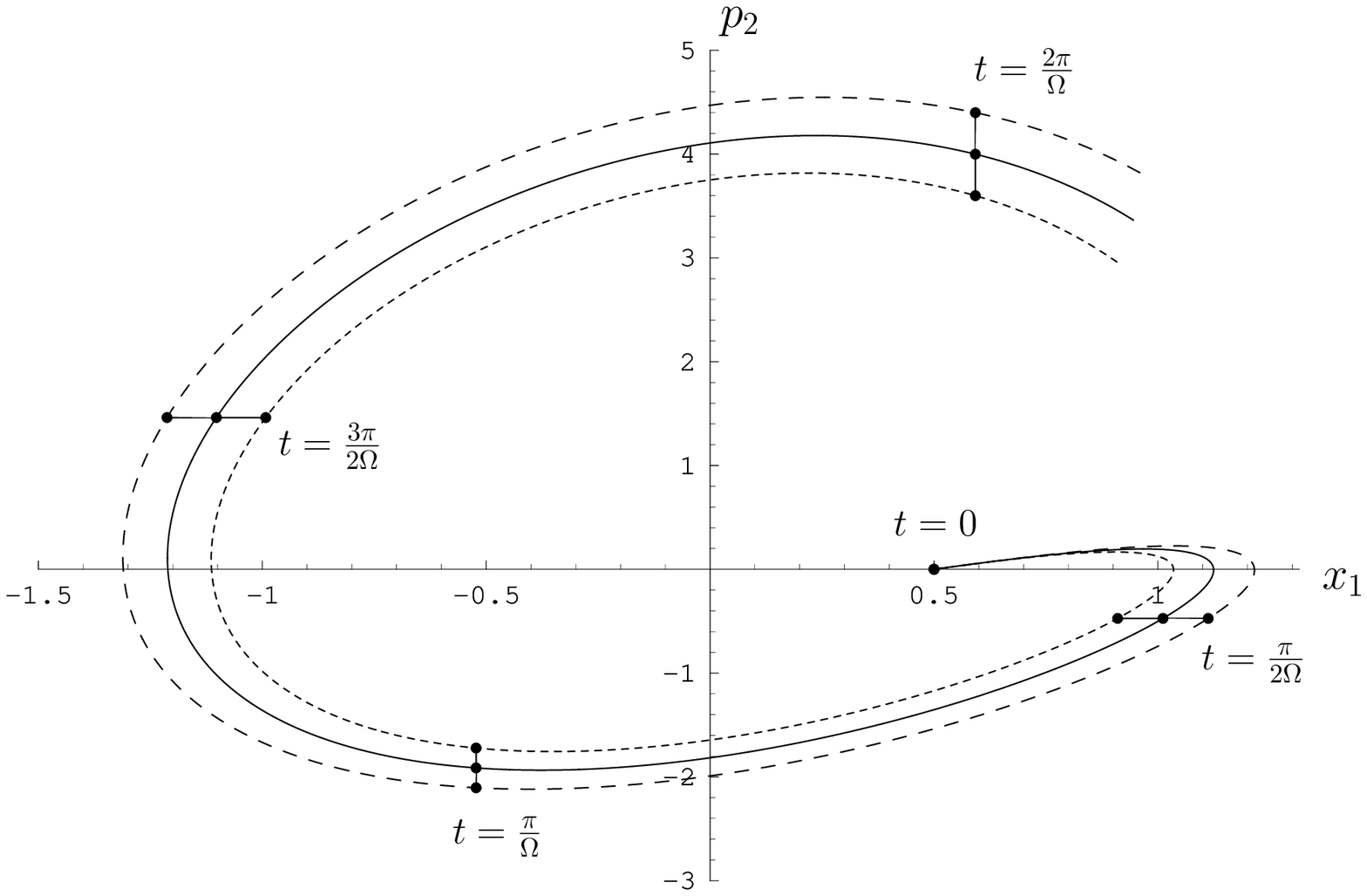}}
\vspace{5mm}
\caption{\em The same plot as Fig.2 in the $(x_1,p_2)$ phase
space.}
\label{fig3}
\end{figure}

\begin{figure}[h]
 \epsfxsize=8.5cm
\centerline{\epsffile{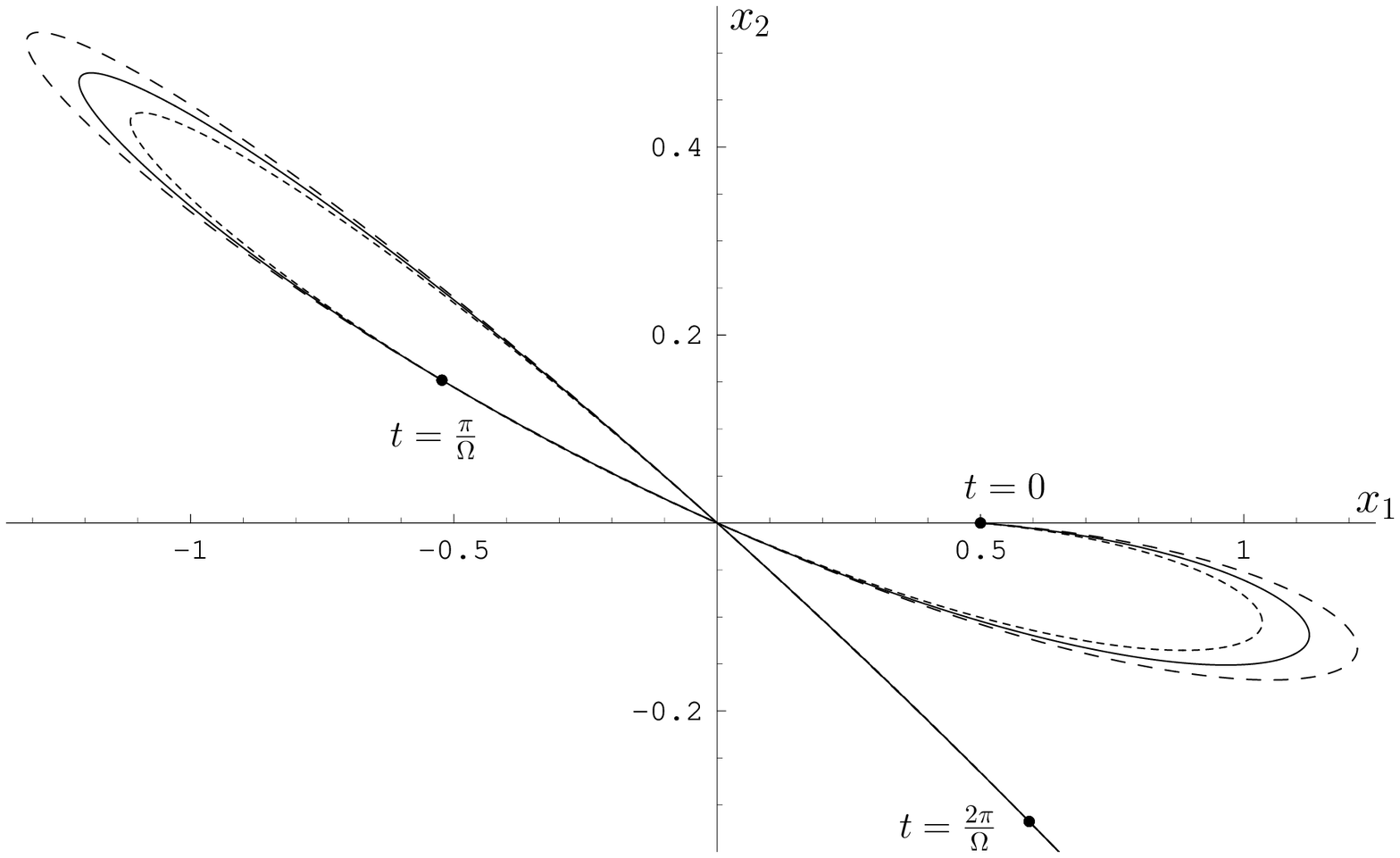}}
\vspace{5mm}
\caption{\em
The corresponding $(x_1,x_2)$ configuration--space orbits. Focal
points occur in ${\bf{x}}_a$ and ${\bf{x}}_b$ at times $t =
\pi/\Omega$ and $t = 2\pi/\Omega$, respectively. Irrespective of the
initial--time momenta, all orbits reach the focal points at
the same time.}
\label{fig4}
\end{figure}

\noindent Let us analyze our particular situation.  Using  the
Hamilton equations of motion we easily get the following solution for
${\bf{z}} \equiv (p_1,p_2,x^1,x^2)$ (to be specific  we work here in
$(x^1,x^2,p_1,p_2)$ phase--space coordinates)
\begin{eqnarray}
{\bf{z}} &=& \exp \left[ (t - t_a)
                 \left(  \begin{array}{cc}
                  \frac{\gamma}{2m}\sigma_1 & -m\Omega^2\sigma_3 \\
                  \frac{1}{m}\sigma_3 &  - \frac{\gamma}{2m} \sigma_1
                   \end{array} \right)
\right]{\bf{z}}_a \nonumber \\
&=& \exp\left[ (t - t_a)
          \left(  \begin{array}{cc}
           0 & -m\Omega^2\sigma_3 \\
                  \frac{1}{m}\sigma_3 & 0
                   \end{array} \right) \right] \nonumber \\
&\times&  \exp \left[ (t - t_a)
                 \left(  \begin{array}{cc}
                  \frac{\gamma}{2m}\sigma_1 & 0 \\
                  0 &  - \frac{\gamma}{2m} \sigma_1
                   \end{array} \right)
\right]{\bf{z}}_a
\, ,
\end{eqnarray}
\noindent and so
\begin{equation}
{\mathbb{S}}_1(t_a,t_b) = \frac{\sin[\Omega (t_b - t_a)]}{\Omega m}
 \ \sigma_3 \, .
\end{equation}
\noindent To get the Morse index for the Berry--Anandan phase
(\ref{gs12}), we simply notice that during the time interval
($0,\tau$) orbits pass one caustic at the conjugate time    $\tau =
\pi/\Omega$, with multiplicity 2 (see
Figs.[\ref{fig2},\ref{fig3},\ref{fig4}]).
The Morse index is $n_{a,b} = 2$ and
the ground state of the 1D l.h.o.  $E_{0}^{lho} =
\hbar \Omega $.  Of course, if $\tau$ is a period for $V$, $2\tau$ is
also a period and any integer  (positive or negative) times $\tau$ is
likewise. It is clear then that the  Morse index based on any such
period should be the same because $V$ (and hence the wave  function) cannot
distinguish whether the fundamental period alone or its integer
multiples are in use.  Indeed, for instance, for the time interval
($0,2\tau$) orbits pass two caustics, at the  conjugate times   $\tau
= \pi/\Omega$ and  $\tau =2\pi/\Omega$, both with  multiplicity 2 (see
Figs.[\ref{fig2},\ref{fig3},\ref{fig4}]).  The Morse index is in such
a case $n_{a,b} = 4$  and we get again $E_{0}^{lho} = \hbar \Omega $.

\noindent

\vspace{3mm}

\noindent We see then that
although our prescription  does not allow to pinpoint
$E_{0}^{lho}$ precisely (the ground states are defined  mod$(\hbar
\Omega)$), there is no way how to bring the usual  fraction $1/2$ into
the result. The factor $2$ is a ``memory'' of the underlying 2D
system. The fact that this may happen is not difficult to
understand. The usual derivations of the ground state energy
hinge either on
the Heisenberg--Weyl algebra  or directly on the Schr{\"o}dinger
equation and thus have only local character while Berry--Anandan phase
is (non--local) global characteristics of a system.  We may thus conclude
that the ground state of the reduced system
can generally change according   to
the global properties of the original underlying system on which the
reduction is performed.

\vspace{3mm}

\noindent Actually the above analysis is still not the whole
story.
In the paper to follow we will show that there is yet
another - not so far considered - contribution to the ground state
energy,  reflecting  the dissipative
nature of the system when working with the $SU(1,1)$ non--unitary
representation. Such a contribution will manifest itself in the form of
an additional phase factor -  ``dissipative'' phase (see also
\cite{BJV1}).

\vspace{5mm}

\section{Conclusions}

\noindent In this paper we have  studied the quantization  of Bateman's
dual system of damped--antidamped harmonic oscillators \cite{BAT,HD1}
by using the Feynman--Hibbs kernel formula.
It has been known for some time that Bateman's system  is
difficult to quantize due to a multitude of  conceptual problems
\cite{HD1,HF1,GV1}. In
order to address some of these problems and  to improve our intuition
for the complications involved, we have found it very convenient that
classical mechanics lies {\em manifestly} at the heart of the
Feynman--Hibbs prescription. Thanks to the fact that Bateman's dual
Hamiltonian (Lagrangian) is quadratic, the kernel is {\em fully}
expressible in terms of the fundamental system of solutions of
classical equations of motion and in addition, it is  independent of
 the choice of such classical solutions.

\vspace{3mm}

\noindent Using the spectral decomposition of the  time--evolution
amplitude  we have been able to calculate the full wave--functions,
fulfilling the time--dependent Scr\"odinger equation. This helped us to
understand the reported  controversy in the quantization of Bateman's
system. We have shown that the above inconsistency
has its origin in two interrelated issues: apparent non--hermiticity
of the Hamiltonian and oddness of $J_2$ under time reversal (and thus
time irreversibility of the Hamiltonian). We have argued that  both
these ``pathologies" are the consequence of a single fact, namely
that what has been  invariably used  in the literature (mostly implicitly)
was the non--unitary irreducible representation of the $SU(1,1)$
dynamic group.  The fact is that from the representation theory it follows
that  there is no   $SU(1,1)$ unitary irreducible representation in
which $J_2$ would have at the same time a real and discrete
spectrum\cite{GL1}. There are then two possible solutions.  We may
work  with the unitary representation of
$SU(1,1)$ -- Bateman's dual system will then be free of pathologies but
not particularly interesting from the physical point of view. The
interesting features, e.g., dissipation along with time
irreversibility,  enter the scene precisely when the non--unitary
irreducible  representation of $SU(1,1)$ is used. To treat the latter
situation mathematically we have to face the non--hermiticity of
$J_2$.  The remedy naturally arises  from the Feynman--Hibbs
prescription and  is based on a redefinition of the inner
product: we have illustrated the
mathematical  and logical consistency of such a procedure. The
relation with existing  results\cite{HD1,HF1,GV1} was established and
the corresponding underlying $SU(1,1)$ coherent state structure of
quantum states was discussed in detail.  The reader may contrast
 our approach with a somewhat more customary
treatment of non--Hermitian quantum  systems by means of resonant or
Gamow states\cite{AM2}.

\vspace{3mm}

\noindent Although the kernel is invariant under the choice of the
fundamental system of classical solutions, this is not the case for
the wave functions. For one--dimensional quadratic systems it has been
argued\cite{DYS1} that various choices of fundamental solutions
correspond merely  to  different unitary transformations.  However,
this shows up not to be correct  in  our case.
In fact, the basic feature of Bateman's dual
system is that states are unitary inequivalent under $SU(1,1)$
symmetry. As a result, the wave functions and thus the
geometric (Pancharatnam) phases are found to depend on  the choice of
the fundamental system of classical solutions in a non--trivial way.
It is also worthwhile stressing that the geometric
phases here obtained are given  only in terms of the parameters of
classical solutions and are thus manifestly  $\hbar$ independent (as
should be  expected  from phases with entirely geometric origin).

\vspace{3mm}

\noindent The crucial observation made in the course of our analysis
is that when we analytically continue the ``azimuthal''  quantum
number $l$ to $\pm 1/2$,  the reduced (or radial) wave functions
formally fulfil the time--dependent Schr{\"o}dinger equation for
the one--dimensional harmonic  oscillator.
We have found that the geometric
phase of the 1D l.h.o. obtained via such a
reduction is not equal to the expected Berry phase \cite{DYS1},
but bears a memory of the classical motion of the original
Bateman dual system.  This ``shadow" of the underlying 2D system originates
from three sources:   overall ground--state dispersions of
$\hat{p}$ and $\hat{x}$ gathered during  the period of evolution and
the Morse index contribution.  It should be noted that the built--in
memory in the geometric phase is not a new idea, having been
used in various contexts as
the Born--Oppenheimer approximation\cite{YA4}
and the dynamic quantum Zeno effect\cite{PF1}.

\vspace{3mm}

\noindent A remarkable feature of the reduction procedure is
that it allows us to find a relation between the ground--state energy
of the 1D l.h.o. obtained after reduction and the  Pancharatnam
phase.    To put some flesh on the bones and to demonstrate the
mechanism  at hand,  we have resorted to  a specific fundamental system
of solutions. The corresponding wave functions then proved to be,
after reduction to those of a simple 1D l.h.o.,  energy eigenstates and
 periodic with the period $\tau$, the latter being connected with
the reduced frequency of the original system.  The reduced geometric
phase - Berry--Anandan phase - was then found to be directly
proportional to the ground--state energy  of the 1D l.h.o..
It was shown that the ground--state energy is controlled
by the Morse index affiliated with  Bateman's  dual system (not with
the 1D l.h.o. itself~!). Finding a Lagrangian
manifold   and following
its evolution in  phase--space, we were able to track down the
number of focal points in the interval $(0,\tau)$ and hence to
identify the Morse index. It turned out that the  ground--state energy
thus acquired is different from the usual $E_{0}=\hbar\Omega/2$.


\vspace{3mm}

%

\noindent Finally, we remark that the reader may find some  resonance
of the method presented here with the results \cite{MCG1,RB1,MVB3} which
suggest that the  quantum mechanical energy spectrum can be determined
from purely classical quantities such as lengths and stability indices
of the periodic orbits alone. Whether or not this formal similarity
can go any further is definitely a challenging question which is being
investigated by the authors\cite{BJV1}.

\section*{Acknowledgements}
\noindent One of us (PJ) would like to thank J.C.~Taylor, T.~Evans,
J.~Tolar and A.~Tanaka for many interesting and illuminating discussions.
This work has been partially supported by the Royal Society and INFN.

\section*{Appendix A}
\noi
It is useful to list some of the expressions of Section II as they look in
$(x,y)$ coordinates.
 The Lagrangian reads:
\begin{eqnarray}
L &=& m\dot{x}\dot{y} + \frac{\gamma}{2}(x \dot{y} - \dot{x} y) -
\kappa xy\nonumber \\ &=& \frac{m}{2} \dot{\bf{x}}\dot{\bf{x}} +
\frac{\gamma}{2}\, \dot{\bf{x}}\wedge {\bf{x}} -
\frac{\kappa}{2}{\bf{x}}{\bf{x}}\, ,
\end{eqnarray}
where  $x^{\alpha}=(x,y)$
with the metric tensor $g_{\alpha \beta} = (\sigma_{1})_{\alpha
\beta}$.  The canonical momenta  are
\begin{equation}
{\bf p} = m {\dot {\bf x}} - \frac{1}{2}\gamma \sigma_{3} {\bf x}\, .
\end{equation}
\noindent and the classical equations of motion can be written as
\begin{equation}\lab{eqxy}
m\, \ddot{\bf{x}}_{cl} + \gamma\sigma_{3}\,\dot{\bf{x}}_{cl} + \kappa
\, {\bf{x}}_{cl} = {\bf{0}}\, .
\end{equation}

\vspace{3mm}

\noindent
Notice   that if ${\bf u}(t)$ is a solution of (\ref{eqxy}) then
$\sigma_{3}{\bf u}(t)$, $\sigma_{1}{\bf u}(-t)$ and $i \sigma_{2} {\bf
u}(-t)$ are also solutions.

\noi The Wronskian for the $(x,y)$ system is
\begin{equation}
W(t) = W(t_{0}) \mbox{exp}\left(-\int_{t_{0}}^{t}dt \,
\mbox{Tr}\left(\frac{\gamma}{m}\sigma_{3}\right) \right)\, ,
\end{equation}
and the action reads
\bea\lab{Sxy}
S_{cl}[{\bf{x}}] &=& \int_{t_{a}}^{t_{b}}dt\,
\left[\frac{m}{2}\left(\frac{d}{dt}(x\dot{y} + \dot{x}y) - x\ddot{y} -
\ddot{x}y \right)\right.\non \\  && \mbox{\hspace{10mm}}+
\left. \frac{\gamma}{2}(x\dot{y} - \dot{x}y )  - \frac{\kappa}{2}xy -
\frac{\kappa}{2}xy \right] \non \\ =&& \frac{m}{2}(x\dot{y} +
\dot{x}y)|_{t_{a}}^{t_{b}}  - \int_{t_{a}}^{t_{b}}dt\,
\frac{\bf{x}}{2}(m\ddot{\bf x}+ \gamma \sigma_{3}\dot{\bf  x} + \kappa
{\bf x} )  \non \\ =&&
\frac{m}{2}[{\bf{x}}_{cl}(t_{b})\dot{\bf{x}}_{cl}(t_{b}) -
{\bf{x}}_{cl}(t_{a})\dot{\bf{x}}_{cl}(t_{a}) ]\, .
\eea
and more explicitly
\begin{eqnarray*}
&&S_{cl}[{\bf x}] = \frac{m}{2D} \left[  -(x_{a}^{1})^{2}\,
\dot{B}_{1}^{2}(t_{a}) -(x_{a}^{2})^{2}\, \dot{B}_{2}^{1}(t_{a})
\right. \non \\ &&
\mbox{\hspace{2mm}}\left. +(x_{b}^{1})^{2}\, \dot{B}_{3}^{2}(t_{b})
+(x_{b}^{2})^{2}\,
\dot{B}_{4}^{1}(t_{b})\right.
\non \\ && \mbox{\hspace{2mm}}\left.-
x_{a}^{1}x_{a}^{2}\, \left(\dot{B}_{1}^{1}(t_{a}) +
\dot{B}_{2}^{2}(t_{a}) \right) + x_{b}^{1}x_{b}^{2}\,
\left(\dot{B}_{4}^{2}(t_{b}) + \dot{B}_{3}^{1}(t_{b})
\right)\right. \non \\ && \mbox{\hspace{2mm}}\left.+
x_{a}^{1}x_{b}^{1}\, \left(\dot{B}_{1}^{2}(t_{b}) -
\dot{B}_{3}^{2}(t_{a}) \right) +  x_{a}^{1}x_{b}^{2}\,
\left(\dot{B}_{1}^{1}(t_{b}) - \dot{B}_{4}^{2}(t_{a})
\right)\right. \non \\  && \mbox{\hspace{2mm}}\left.-
x_{a}^{2}x_{b}^{2}\, \left(\dot{B}_{4}^{1}(t_{a}) -
\dot{B}_{2}^{1}(t_{b}) \right) + x_{a}^{2}x_{b}^{1}\,
\left(\dot{B}_{2}^{2}(t_{b}) - \dot{B}_{3}^{1}(t_{a}) \right)
\right]\, .  \,
\end{eqnarray*}

\noi Finally, the fluctuation factor in
$(x,y)$ coordinates is
\bea &&F[t_{a}, t_{b}] = \frac{m}{4\pi \hbar D}\left[-\left(
\dot{B}^{2}_{1}(t_{b})\dot{B}^{2}_{2}(t_{b}) -
\dot{B}^{2}_{3}(t_{a})\dot{B}^{2}_{2}(t_{b}) \right.\right.\non \\
&&\mbox{\hspace{6mm}}\left.\left. -
\dot{B}^{2}_{1}(t_{b})\dot{B}^{1}_{3}(t_{a}) +
\dot{B}^{2}_{3}(t_{a})\dot{B}^{1}_{3}(t_{a}) +
\dot{B}^{1}_{4}(t_{a})\dot{B}^{1}_{1}(t_{b}) \right.\right.\non  \\
&&\mbox{\hspace{6mm}} \left.\left.-
\dot{B}^{1}_{4}(t_{a})\dot{B}^{2}_{4}(t_{a}) -
\dot{B}^{1}_{2}(t_{b})\dot{B}^{1}_{1}(t_{b}) +
\dot{B}^{1}_{2}(t_{b})\dot{B}^{2}_{4}(t_{a})
\right)\right]^{\frac{1}{2}} \non  \\  &&\mbox{\hspace{12mm}} =
\frac{m}{2\pi \hbar} \sqrt{\frac{W}{D}} \, .
\eea

\noindent Notice that under
transformation $(x_{1},x_{2}) \rightarrow (x,y)$ both $W$ and $D$ do
not change.

\section*{Appendix B}
\noi We prove here
that  both $S_{cl}[{\bf x}]$ and ${\bf x}_{cl}(t)$ are independent of
the choice of the fundamental system. The proof can be done in two
steps. Firstly we show that both $S_{cl}[{\bf x}]$ and ${\bf x}_{cl}$
do not depend on the scaling of ${\bf u}_{i}$ and ${\bf
v}_{i}$. Indeed, if we rescale, for example ${\bf u}_{1} \rightarrow
\alpha {\bf u}_{1}$, then
\begin{eqnarray*}
&&U_{a} \rightarrow \alpha U_{a}\qquad ,\qquad V_{b} \rightarrow V_{b}\,
,\\ &&D_{1} \rightarrow D_{1}\qquad , \qquad  D_{2} \rightarrow \alpha
D_{2}\, ,\\  &&D_{3} \rightarrow \alpha D_{3}\qquad ,\qquad  D_{4}
\rightarrow \alpha D_{4}\, .
\end{eqnarray*}
\noindent Analogous relations are valid for other vectors. It is now
simple to see that both $S_{cl}$ and ${\bf x}_{cl}$ remain unchanged
under such a rescaling.

\vspace{3mm}

\noindent Secondly, we show that $S_{cl}$ and ${\bf x}_{cl}$ remain
unchanged under the substitution
\begin{equation}
{\bf u}_{1} \rightarrow {\bf u}_{1} + \alpha {\bf u}_{2} + \beta {\bf
v}_{1} + \gamma {\bf v}_{2} \, ,
\label{sub1}
\end{equation}
\noindent where $\alpha, \beta, \gamma$ are arbitrary real constants.
The previous substitution is possible to achieve, for example, in
successive steps:
\begin{eqnarray*}
&&{\bf u}_{1} \rightarrow {\bf u}_{1} +\beta {\bf v}_{1}\, ,\\  &&{\bf
v}_{1} \rightarrow {\bf v}_{1} + \frac{\gamma}{\beta} {\bf v}_{2}\,
,\\ &&{\bf v}_{2} \rightarrow {\bf v}_{2} + \frac{\alpha}{\gamma} {\bf
u}_{2}\, .
\end{eqnarray*}
\noindent It may be directly seen that both $S_{cl}$ and ${\bf
x}_{cl}$ are invariant under each of the former substitutions and so
they are invariant with respect to (\ref{sub1}), too. To see it more
explicitly let us perform, for instance, the substitution ${\bf u}_{1}
\rightarrow {\bf u}_{1} + \beta {\bf v}_{1}$, then
\begin{eqnarray*}
&&U_{a} \rightarrow U_{a}\qquad ,\qquad V_{b} \rightarrow V_{b}\, ,\\
&&D_{1} \rightarrow D_{1}\qquad ,\qquad D_{2} \rightarrow D_{2}\, ,\\
&& D_{4} \rightarrow
D_{4}\qquad ,\qquad D_{3} \rightarrow D_{3} - \beta D_{1}\, .
\end{eqnarray*}
\noindent Plugging the previous substitution into (\ref{cr1}), we get
that ${\bf x}_{cl}(t) \rightarrow {\bf x}_{cl}(t)$. The same is true
for $S_{cl}$ as it might be directly seen from relations
(\ref{e7}) and (\ref{Sxy}).

\vspace{3mm}

\noindent The previous two observations therefore lead us to the
conclusion that $S_{cl}$ does not depend on the particular choice of the
fundamental system of solutions (expressible as a
linear combination of ${\bf u}_{i}$ and ${\bf v}_{i}$).

\section*{Appendix C}

\noindent
We derive here some relations  used in Section IIIA.
As we have mentioned in Subsection IIB,
having one solution, say for $(x_{1},x_{2})$ coordinates, we can get
another one if we multiply the original one by $\sigma_{3}$. So namely
if one has the fundamental system of solutions
$({\bf u}_{1}, {\bf u}_{2}, {\bf
v}_{1}, {\bf v}_{2})$, one can generate another fundamental system
$(\sigma_{3}{\bf u}_{1}, \sigma_{3} {\bf u}_{2}, \sigma_{3} {\bf
v}_{1}, \sigma_{3} {\bf v}_{2})$ (it is simple to see that this is
indeed a fundamental system by looking at the Wronskian).
Identical reasonings
as in the Subsection IID will lead us to the result
\bea &&{\bf x}_{cl}(t) = \frac{1}{D}\left[ x_{a}^{1} \left(
\begin{array}{l} B_{2}^{2}(t)\\ B_{2}^{1}(t)
\end{array} \right)
+ x_{a}^{2} \left( \begin{array}{l} B_{1}^{2}(t)\\ B_{1}^{1}(t)
\end{array} \right) \right. \non \\
&&\mbox{\hspace{1.4cm}}+ \, \left.   x_{b}^{1} \left( \begin{array}{l}
B_{4}^{2}(t)\\ B_{4}^{1}(t)
\end{array} \right)
+ x_{b}^{2} \left( \begin{array}{l} B_{3}^{2}(t)\\ B_{3}^{1}(t)
\end{array} \right) \right]\, .
\label{eq45}
\eea
\noindent Comparing (\ref{eq45}) with (\ref{eq44}) we get the
following useful identities:
\bea B_{1}^{1}(t) &=& B^{2}_{2}(t)\qquad , \qquad B_{1}^{2}(t) =
B^{1}_{2}(t)\, ,\non \\  B_{3}^{1}(t) &=& B^{2}_{4}(t)\qquad ,\qquad
B_{3}^{2}(t) = B^{1}_{4}(t)\, .
\label{id1}
\eea
\noindent Similar analysis may be done with the fundamental systems
$(\sigma_{3}{\bf u}_{1}(-t), \sigma_{3}{\bf u}_{2}(-t), \sigma_{3}{\bf
v}_{1}(-t), \sigma_{3}{\bf v}_{2}(-t))$ and $(i\sigma_{2}{\bf
u}_{1}(-t), i\sigma_{2}{\bf u}_{2}(-t), i\sigma_{2}{\bf v}_{1}(-t),
i\sigma_{2}{\bf v}_{2}(-t))$.  In this case it can be directly checked
that we have
\begin{equation}
B_{i}^{j}(t) = (-1)^{i+j}\,B_{i}^{j}(-t)\, .
\end{equation}
\noindent From the definition (\ref{det}) and relations (\ref{id1}) we
may observe that
\bea \frac{d}{d t_{a}}D &=& 2 {\dot B}_{1}^{1}(t_{a})\quad , \quad
\frac{d}{d t_{b}}D = 2 {\dot B}_{3}^{1}(t_{b})\, , \eea
\noindent so namely
\begin{equation}
\; \; \; \; \; \; 2\,\frac{{\dot B}^{1}_{1}(t_{a})}{D} =
\frac{d}{dt_{a}}  \mbox{ln} D = {\mbox Tr}\left({\bf D}^{-1}
\frac{d}{dt_{a}} {\bf D}\right)\, ,
\label{B1}
\end{equation}
\begin{equation}
2\,\frac{{\dot B}^{1}_{3}(t_{b})}{D} = \frac{d}{dt_{b}} \mbox{ln} D  =
{\mbox Tr} \left({\bf D}^{-1} \frac{d}{dt_{b}} {\bf D}\right)\, .
\label{Bs}
\end{equation}
\noindent Here $D = \mbox{det}{\bf D}$. Because $D(t_{a}, t_{b})
=D(t_{b} - t_{a})$ it fulfills the equation
\begin{displaymath}
\frac{d D(t_{a},t_{b})}{d t_{a}} + \frac{d D(t_{a},t_{b})}{d t_{b}} =
0\, ,
\end{displaymath}
\noindent and so we have the identity
\begin{equation}
{\dot B}^{1}_{1}(t_{a}) = - {\dot B}^{1}_{3}(t_{b})\, .
\end{equation}
%

\section*{Appendix D}

\noindent We prove here the relation (\ref{alpha1}). For
this  purpose it is simpler to work in the $(x,y)$ coordinates. The
kernel can be then constructed in analogous way as in $(x_{1},x_{2})$
coordinate.  A simple calculation shows that
\begin{eqnarray}
S_{cl}[r,u] &=& \frac{m}{2D} \left[ - \frac{r_{a}^{2}}{2}\,
\frac{dD}{d t_{a}} + \frac{r_{b}^{2}}{2}\, \frac{dD}{d t_{b}}
\right.\nonumber \\ &+& \left. r_{a} r_{b} \left( e^{u_{a} - u_{b}} \,
{\dot B}^{1}_{1}(t_{b}) + e^{u_{b} - u_{a}}\, {\dot B}^{2}_{2}(t_{b})
\right)\right] \, ,
\end{eqnarray}
\noindent and that
\begin{displaymath}
{\dot B}_{1}^{1}(t_{b}){\dot B}_{2}^{2}(t_{b}) = WD\, .
\end{displaymath}
\noindent The latter allows to identify
\begin{displaymath}
\frac{{\dot B}_{1}^{1}(t_{b})}{D} = \sqrt{\frac{W}{D}} \, e^{\alpha}
\, ; \;\;\;\;\;\; \frac{{\dot B}_{2}^{2}(t_{b})}{D} =
\sqrt{\frac{W}{D}} \ e^{-\alpha} \, .
\end{displaymath}
\noindent This identification fixes $\alpha$ modulo $i\pi$ and leads
to the equation
\begin{equation}
\alpha = \frac{1}{2}\, \mbox{ln}\left(\frac{{\dot
B}_{1}^{1}(t_{b})}{{\dot B}_{2}^{2}(t_{b})} \right)\, .
\label{a11}
\end{equation}

\noindent In addition, from the symmetry reasonings result the
following useful relations
\begin{eqnarray*}
{\dot B}_{1}^{1}(t_{b}) &=& - {\dot B}_{4}^{2}(t_{a}) \quad , \quad
{\dot B}_{2}^{2}(t_{b}) = - {\dot B}_{3}^{1}(t_{a})\, .
\end{eqnarray*}
\noindent Let us now consider ${\dot B}^{1}_{1}(t_{b})$. Its explicit
structure reads
\begin{equation}
{\dot B}^{1}_{1}(t_{b}) = \left| \begin{array}{llll} {\dot
u}_{1}^{1}(t_{b}) & {\dot u}_{2}^{1}(t_{b}) & {\dot v}_{1}^{1}(t_{b})
& {\dot v}_{2}^{1}(t_{b}) \\ u_{1}^{2}(t_{a}) & u_{2}^{2}(t_{a}) &
v_{1}^{2}(t_{a}) & v_{2}^{2}(t_{a}) \\ {\bf u}_{1}(t_{b}) & {\bf
u}_{2}(t_{b}) & {\bf v}_{1}(t_{b}) & {\bf v}_{2}(t_{b})
\end{array} \right|\, .
\end{equation}
\noindent Rules for differentiation of determinants tell that ${\dot
B}^{1}_{1}(t_{b})$ fulfills the equation
\begin{equation}
\frac{d^{2}}{d t_{a}^{2}} {\dot     B}^{1}_{1}(t_{b}) -
\frac{\gamma}{m}\, \frac{d}{dt_{a}} {\dot     B}^{1}_{1}(t_{b}) +
\frac{\kappa}{m}\, {\dot     B}^{1}_{1}(t_{b}) = 0\, ,
\label{a1}
\end{equation}
\noindent with the boundary condition ${\dot B}^{1}_{1}(t_{b})|_{t_{a}
= t_{b}} = 0$. The general solution of (\ref{a1}) reads
\begin{equation}
{\dot B}^{1}_{1}(t_{b}) = e^{\Gamma(t_{a})}\left(e^{i\Omega t_{a}}
f(t_{b}) + e^{-i\Omega t_{a}} {\tilde f}(t_{b}) \right)\, .
\end{equation}
\noindent Here $f$ and $\tilde f$ are some functions of
$t_{b}$. Applying the boundary condition we get
\begin{equation}
{\dot B}^{1}_{1}(t_{b}) = C\, e^{-\Gamma(t_b - t_{a})}\,
\mbox{sin}\Omega(t_{b} - t_{a}) \, ,
\label{a3}
\end{equation}
\noindent with $C$ being a constant. The result is clearly the only
one   which is compatible with the differential equation for ${\dot
B}^{2}_{4}(t_{a})$ ($= - {\dot B}^{1}_{1}(t_{b})$):
\begin{equation}
\frac{d^{2}}{d t_{b}^{2}} {\dot B}^{2}_{4}(t_{a}) + \frac{\gamma}{m}
\, \frac{d}{dt_{b}} {\dot B}^{2}_{4}(t_{a}) + \frac{\kappa}{m} \,
{\dot B}^{2}_{4}(t_{a}) = 0 \, ,
\end{equation}
\noindent fulfilling the boundary condition ${\dot
B}^{2}_{4}(t_{a})|_{t_{b} = t_{a}} = 0$.

\vspace{3mm}

\noindent The same reasonings can be now applied on ${\dot
B}^{2}_{2}(t_{b})$. The latter fulfills the differential equation
\begin{equation}
\frac{d^{2}}{d t_{a}^{2}} {\dot B}^{2}_{2}(t_{b}) + \frac{\gamma}{m}\,
\frac{d}{dt_{a}} {\dot B}^{2}_{2}(t_{b}) + \frac{\kappa}{m}\, {\dot
B}^{2}_{2}(t_{b}) = 0\, ,
\label{a4}
\end{equation}
\noindent with the boundary condition ${\dot B}^{2}_{2}(t_{b})|_{t_{a}
= t_{b}} = 0$. The solution is
\begin{equation}
{\dot B}^{2}_{2}(t_{b}) = {\tilde C}\, e^{\Gamma(t_b - t_{a})}\,
\mbox{sin}\Omega(t_{b} - t_{a}) \, ,
\label{a33}
\end{equation}
\noindent with ${\tilde C}$ being a constant. The result is the only
one which is compatible with the differential equation for ${\dot
B}^{1}_{3}(t_{a})$ ($= - {\dot B}^{2}_{2}(t_{b})$):
\begin{equation}
\frac{d^{2}}{d t_{b}^{2}} {\dot B}^{1}_{3}(t_{a}) - \frac{\gamma}{m}
\, \frac{d}{dt_{b}} {\dot B}^{1}_{3}(t_{a}) + \frac{\kappa}{m} \,
{\dot B}^{1}_{3}(t_{a}) = 0 \, ,
\end{equation}
\noindent fulfilling the boundary condition ${\dot
B}^{1}_{3}(t_{a})|_{t_{b} = t_{a}} = 0$.

\noi Gathering the results
(\ref{a11}), (\ref{a3}) and (\ref{a33}) together we obtain
\begin{equation}
\alpha(t_{a},t_{b}) = \Gamma(t_{a} - t_{b}) + {\mbox{ln}}C -
{\mbox{ln}}{\tilde C}\, ,
\end{equation}
\noindent modulo $i\pi$. Using l'Hopital rule one may persuade oneself
that $\lim_{t_{a} \rightarrow t_{b}}\left|{\dot B}^{2}_{2}(t_{b}) /
{\dot B}^{1}_{1}(t_{b}) \right| = \left| C/{\tilde C} \right| =
1$. Thus, $\mbox{ln}C - \mbox{ln}{\tilde C}$ is either zero or purely
imaginary.

\section*{Appendix E}

\noindent Let us first emphasize that the formulation of a
time reversal transformation must avoid using properties of the forces
or interactions that determine the dynamics, because it is
the transformation properties of the  dynamic equations which we seek to
determine.  The latter is the crux often overlooked by many authors.  Since
the kinematics  are those properties of the motion that are
independent of the dynamics,  we
require that the ``admissible'' time--reversal transformation should be
formulated in kinematic terms.  This means that the
 time--reversal transformation  must be
consistent with the algebraic structure of the operators representing
the (kinematic) observables  and that in the absence of forces or
interactions (i.e., in the absence of causal effects), the dynamic
equations must be left invariant.

\vspace{3mm}

\noindent In our case the kinematic observables may be taken to be
$x^{\alpha}$ and $P^{\alpha}$ (note that ${\bf{P}}= m {\dot {\bf x}}$
 are the kinetic
momenta and not the full canonical momenta (\ref{cm12})).
Working in $(x_1,x_2)$ coordinates,
the algebraic structure is then determined
by the Heisenberg--Weyl group
\begin{equation}
[x^{\alpha}, P^{\beta}] = i\hbar (\sigma_3)^{\alpha \beta}\,,
\;\;\;\;\;\; [{\bold{x}}, \sigma_3 ] = [{\bold{P}}, \sigma_3] =0\, ,
\label{HW1}
\end{equation}
\noindent
On the other hand, if $\hat{H}_{0}$ is the Hamiltonian
in the absence of interaction ($\gamma = 0$), the dynamic
(Schr{\"o}dinger) equation
\begin{equation}
i\hbar \, \frac{d}{dt}|\psi(t)\rangle = \hat{H}_{0}
|\psi(t)\rangle \, , \label{TR1}
\end{equation}
\noindent must transform under time reversal $\cal{T}$ into
\begin{equation}
i\hbar \, \frac{d}{dt'}|\psi'(t')\rangle = \hat{H}_{0}
|\psi'(t')\rangle \, , \label{TR2}
\end{equation}
\noindent where $t' = -t$.
Applying  $\cal{T}$ to both sides of (\ref{TR1}) we obtain
\begin{equation}
{\cal{T}}i{\cal{T}}^{-1}\hbar \, \frac{d}{dt}|\psi'(t')\rangle =
{\cal{T}}\hat{H}_{0}{\cal{T}}^{-1} |\psi'(t')\rangle \, .
\label{TR3}
\end{equation}
\noindent Comparing (\ref{TR3}) with (\ref{TR2}) and using the
requirement that ${\hat{H}}_0$ is invariant under time reversal we obtain
the relation: ${\cal{T}}i{\cal{T}}^{-1} = -i$. Invoking
 Wigner's theorem\cite{Wi1}, the latter implies that $\cal{T}$ must be
antiunitary and thus
may be written as $\cal{T} = UK$ with ${\cal{K}}$ being the
complex conjugation operator and ${\cal{U}}$ being some unitary
operator.

\vspace{3mm}

\noindent At the same time, in accordance with the classical conditions,
time reversal requires that
\begin{equation}
{\bold{x}}_\TR = {\mathbb{B}} {\bold{x}}, \;\;\;\;\; {\bold{P}}_\TR =
-{\mathbb{B}} {\bold{P}}\, ,
\end{equation}
\noindent
The matrix
${\mathbb{B}}$ is supposed to leave $\hat{H}_0$ invariant under the
time reversal.  This means that ${\mathbb{B}}^t (\sigma_3){\mathbb{B}}
= \sigma_3 $ and  ${\mathbb{B}}^2 = 1$  (i.e., time reversal when
repeated must restore the original situation). So ${\mathbb{B}}$ must
be part of a  discrete (two--element) subgroup of  $O(1,1;R)$
(i.e., the real Lorentz group in the plane).
It is well known that the only matrices
fulfilling the above conditions are  $\pm 1$ and $\pm \sigma_3$.

\vspace{3mm}

\noindent To decide the form of ${\mathbb{B}}$ we use the fact
that ${\hat{H}}_0$ is invariant under $O(1,1;R)$. This
means that ${\bf{x}}$ and ${\bf{P}}$ transform under $O(1,1;R)$ in the
usual manner, i.e.,
\begin{eqnarray}
&&{\bf{x}}' = U(\varepsilon){\bf{x}} U^{-1}(\varepsilon)
= {\mathbb{G}}(\varepsilon)
{\bf{x}}\, ,\nonumber \\
&&{\bf{P}}' = U(\varepsilon){\bf{P}} U^{-1}(\varepsilon) =
{\mathbb{G}}(\varepsilon)
{\bf{P}}\, ,
\end{eqnarray}
\noindent where $U(\varepsilon)$ is a unitary representation of
$O(1,1;R)$ in the state space   and ${\mathbb{G}}(\varepsilon)$ is
an element of $O(1,1;R)$ in 2--dimensional vector space. As a
result, the following relation must hold for any ${\mathbb{G}}\in
O(1,1;R)$:
\begin{eqnarray}
U{\cal{T}}{\bf{x}} {\cal{T}}^{-1} U^{-1} =
U{\bf{x}}_\TR U^{-1}= {\mathbb{G}}{\bf{x}}_\TR = {\mathbb{G}}
{\mathbb{B}}{\bf{x}}\, ,
\label{pj66}
\end{eqnarray}
\noindent However, the very same relation may be recast in a slightly
different form, namely
\begin{eqnarray}
U{\cal{T}}{\bf{x}} {\cal{T}}^{-1} U^{-1}
&=& U {\mathbb{B}} {\bf{x}}  U^{-1}\nonumber \\
&=&
U {\mathbb{B}}  U^{-1}
{\mathbb{G}}{\bf{x}} =
{\mathbb{G}}^{t} {\mathbb{B}} {\mathbb{G}}^2
{\bf{x}} \, .
\label{pj24}
\end{eqnarray}
\noindent Comparing both (\ref{pj66}) and (\ref{pj24}) we get that
${\mathbb{G}}{\mathbb{B}} ={\mathbb{G}}^{t} {\mathbb{B}}
{\mathbb{G}}^{2}$. As this must be true for all
${\mathbb{G}}\in O(1,1;R)$, we can choose
\begin{equation}
{\mathbb{G}}(\varepsilon) = \exp\left(\varepsilon \sigma_1 \right)
= \cosh(\varepsilon) + \sinh(\varepsilon) \sigma_1  \, .
\end{equation}
\noindent It is then obvious that the
case ${\mathbb{B}} = \pm 1$ is ruled out
and we are left with ${\mathbb{B}} = \pm \sigma_3$.  However, the
``$+$'' sign is the only
plausible one.  This is because the signature of  the time reversal
should be preserved under continuous  change of coordinates and so
namely when  we shrink the $x_2$ coordinate into the
origin  (i.e., perform a
dimensional reduction) $x_1$ coordinate  must  behave under time
reversal as in ordinary 1D l.h.o..

\vspace{3mm}

\noindent As a upshot of the performed analysis
we have the following transformations:
\begin{eqnarray}
&& {\bold{x}} \rightarrow {\bold{x}}_\TR  =
{\cal{T}}{\bold{x}}{\cal{T}}^{-1}
= \sigma_3 {\bold{x}}\, , \nonumber \\
&& {\bold{p}} \rightarrow {\bold{p}}_\TR ={\mathcal{T}} {\bold{p}}
{\mathcal{T}}^{-1}  = - \ \sigma_3
{\bold{p}}\, , \nonumber \\
&& r \rightarrow r_\TR ={\mathcal{T}} r {\mathcal{T}}^{-1} =
r\, , \nonumber \\
&& u \rightarrow u_\TR ={\mathcal{T}} u {\mathcal{T}}^{-1}  = -u\, ,
\label{pj678}
\end{eqnarray}
\noindent(here ${\bold{p}}$ are full, i.e.,
canonical momenta). Similarly, we find that
\begin{eqnarray}
&& {\cal{T}}{A}{\cal{T}}^{-1} =
- \ A \, , \qquad \;{\cal{T}}{B}{\cal{T}}^{-1} =
 B \, , \nonumber \\
&& {\cal{T}}J_{+}{\cal{T}}^{-1} = - J_{+}\, ,
\quad \; \,\,{\cal{T}}J_{-}{\cal{T}}^{-1} = - J_{-}\, ,\nonumber \\
&& {\cal{T}}J_{1}{\cal{T}}^{-1} = - J_{1}\, ,
\qquad \, {\cal{T}}J_{3}{\cal{T}}^{-1} = J_{3}\, ,
\label{pj679}
\end{eqnarray}
\noindent  and the time--reversed
commutation relations
\begin{equation}
[x^{\alpha}_\TR, p^{\beta}_\TR] = -i\hbar (\sigma_3)^{\alpha \beta}\,,
\;\;\;\; [{\bold{x}}_\TR, \sigma_3 ] = [{\bold{p}}_\TR, \sigma_3] =0\, .
\label{HW3}
\end{equation}
\noindent Transformation rules (\ref{pj678}) or (\ref{pj679}) assert that
\begin{equation}
{\cal{T}} {\cal{C}}{\cal{T}}^{-1} = {\cal{C}}\, , \;\;\; \; {\cal{T}}
J_{2} {\cal{T}}^{-1} =  J_{2}\, .
\label{TI1}
\end{equation}
\noindent We thus finally  arrive at the conclusion that
${\cal{T}}{\hat H}{\cal{T}}^{-1}= {\hat H}$. The latter is not
actually compatible with the time reversal presented  in
Ref.\cite{HD1}, where ${\mathbb{B}} =1$ was incorrectly assumed.

\section*{Appendix F}

\noi Using the fact that $SU(1,1)$ ladder operators are $J_+$ and
$J_-$  and the $SU(1,1)$ vacuum state is the state $|j, |j|\rangle$
(i.e., $J_-\,|j, |j|\rangle =0$), we may write\cite{APE1}:
\begin{eqnarray}
&&|\psi^s_{n,l}\rangle  = \left( \frac{1}{\sqrt{2}} \right)^{2|j|
+1}\, \sqrt{\frac{(2|j|)!(m-|j|)!}{(m+|j|)!}} \nonumber \\  &&
\mbox{\hspace{1cm}} \times \;\; L^{2|j|}_{m-|j|} \left( -
\frac{J_+}{2} \right) \, | 1 \rangle \rangle  \, , \nonumber \\ &&
\nonumber \\ && j = n + \mbox{$\frac{l}{2}$} + \mbox{$\frac{1}{2}$}\,
; \;\;\; m = \mbox{$\frac{l}{2}$} - \mbox{$\frac{1}{2}$}\, .
\label{pp21}
\end{eqnarray}

\noindent Here $|z  \rangle \rangle = \mbox{exp}(z J_+) \ |j,
|j|\rangle $  is the (unnormalized) coherent state of $SU(1,1)$
\cite{APE1} (which can be identified with the Gelfand--Neimark
$z$--basis\cite{GN1}).

\vspace{3mm}

\noindent In deriving Eq.(\ref{pp21}) we
have employed (\ref{psis22}) together with the ``annihilation''
and ``creation'' relations
\begin{eqnarray}
(J_-)^{k}|j,m\rangle &=& a_{j,m,k}\, |j, m-k\rangle\, ,
\;\;\; m-k \geq |j| \nonumber \\
                     &=& 0 \, , \;\;\; \; \; \; \; \; \; \; \;\;\;\;\;\;
\;\;\;\;\;\;\;\;\;\;
m-k \leq |j|\, , \nonumber \\
(J_+)^{k}|j,m\rangle &=& b_{j,m,k}\, |j, m+k\rangle\, ,
\end{eqnarray}
\noindent where
\begin{eqnarray*}
a_{j,m,k} &=& \sqrt{\frac{(m-j)! (m+j)!}
{(m-j -k)!(m+j -k)!}} \, , \nonumber \\
b_{j,m,k} &=& \sqrt{\frac{(m+k+j)! (m+k-j)!}
{(m+j)!(m-j)!}} \, .
\end{eqnarray*}

\noindent Note that $(J_+ J_-) \ |j,m\rangle = \left( m^2 - j^2 \right)
|j,m\rangle $.
Form (\ref{pp21}) suggests that the state $|\psi^s_{n,l}\rangle$
can be alternatively interpreted as an excited $SU(1,1)$
coherent state $|1\rangle \rangle$. Because $L_{n}^{\alpha}(x)$
is a polynomial of $n$-th order in $x$,
the relation (\ref{pp21}) asserts that there
is up to
$(m-|j|)$ new ``$SU(1,1)$ excitations'' condensed into the coherent state
$|1\rangle \rangle$.


\end{document}